\documentclass{aastex631}

\newcommand{\detectionthreshold}{7}

\newcommand{\boxcarwindowsize}{$500\,{\rm s}$}
\newcommand{\numvortices}{309}
\newcommand{\discardedvortices}{22}
\newcommand{\vorticesabovelowercutoff}{111}
\newcommand{\vorticesaboveuppercutoff}{19}
\newcommand{\numRDSexcursions}{75}

\accepted{2021 Dec 20}

\submitjournal{PSJ}

\shorttitle{Vortices and Dust Devils As Observed by MEDA on Mars 2020}
\shortauthors{Jackson}
\graphicspath{{./}{figures/}}

\begin{document}

\title{Vortices and Dust Devils As Observed by the MEDA Instruments onboard Mars 2020 Perseverance Rover}

\author[0000-0002-9495-9700]{Brian Jackson}
\affiliation{Department of Physics\\ 
Boise State University\\ 
1910 University Drive, Boise ID 83725-1570 USA}



\begin{abstract}
An important and perhaps dominant source of dust in the martian atmosphere, dust devils play a key role in Mars' climate. Datasets from previous landed missions have revealed dust devil activity, constrained their structures, and elucidated their dust-lifting capacities. However, each landing site and observational season exhibits unique meteorological properties that shape dust devil activity and help illuminate their dependence on ambient conditions. The recent release of data from the Mars Environmental Dynamics Analyzer (MEDA) instrument suite onboard the Mars 2020 Perseverance rover promises a new treasure-trove for dust devil studies. In this study, we sift the time-series from MEDA's Pressure Sensor (PS) and Radiative and Dust Sensors (RDS) to look for the signals of passing vortices and dust devils. We detected \numvortices\ vortex encounters over the mission's first 89 sols. Consistent with predictions, these encounter rates exceed InSight and Curiosity's encounter rates. The RDS time-series also allows us to assess whether a passing vortex is likely to be dusty (and therefore is a true dust devil) or dustless. We find that about one quarter of vortices show signs of dust-lofting, although unfavorable encounter geometries may have prevented us from detecting dust for other vortices. In addition to these results, we discuss prospects for vortex studies as additional data from Mars 2020 are processed and made available. 
\end{abstract}



\section{Introduction} \label{sec:Introduction}

The Mars 2020 Perseverance rover landed on 2021 February 18 ($L_{\rm s} = 5.6^\circ$ -- \url{http://www.tinyurl.com/MarsClock}) at the Octavia E.~Butler Landing site within Jezero Crater on Mars ($18.4447^\circ$ N, $77.4508^\circ$ E). The primary goals of the mission are to seek signs of extant and extinct life and collect rock and soil samples for a future return to Earth by acquiring imaging, spectroscopy, and other measurements to characterize Martian soils, rocks, atmosphere, and other aspects of the environment \citep{2020SSRv..216..142F}. To address these goals, the rover carries seven scientific instruments, as well as a sample acquisition and caching system. 

These instruments include the Mars Environmental Dynamics Analyzer (MEDA) suite consisting of sensors to measure environmental variables -- air pressure and temperature (the pressure and temperature sensors, PS and ATS respectively), up/downward-welling radiation and dust optical depth (via the Radiation and Dust Sensor RDS), wind speed and direction (wind sensors 1 and 2, WS1 and 2), relative humidity (via the humidity sensor HS), and ground temperature (via the Thermal Infrared Sensor TIRS). This combination of powerful, accurate, and precise instrumentation will enable novel investigations of atmospheric processes on Mars, ranging from estimation of the near-surface radiation budget on sub-diurnal and longer timescales to exploration of the role of dust in thermal forcing to investigation of the wind stress thresholds for driving aeolian transport \citep{2021SSRv..217...48R}.

Small-scale, dry, and dust-laden convective vortices, dust devils act as a key if ephemeral aeolian transport mechanism on the surface of Mars, lofting a significant fraction of the dust in the martian atmosphere \citep{2016SSRv..203...89F}. Observations of martian dust devils go back to the Viking mission \citep{1985Sci...230..175T, 2003Icar..163...78R}, and they appear frequently in imagery from landed and orbiting spacecraft \citep{2016SSRv..203...39M, 2015Icar..260..246F}. As a boundary layer phenomenon, they also register in meteorological datasets collected both on Mars and the Earth. These signals come in the form of short-lived (few to tens of seconds), negative pressure excursions ($\Delta P \lesssim 1\%$ of the ambient pressure), accompanied by rapid changes in wind speed and direction \citep{2021Icar..35914207K}. 

However, pressure and wind excursions do not suffice to distinguish dust devils, which are vortices carrying dust, from dustless vortices, which are governed by much the same physics \citep{2016Icar..278..180S}. Indeed, the precise conditions that allow a dustless vortex to become a dust devil are not clear but likely depend on the availability of dust and the vortex wind speeds. Instrumentation measuring solar insolation, alongside pressure and winds, can be used to determine whether a passing vortex is dust-laden or not: a dusty vortex can register either a dip in insolation (if the dust devil's shadow passes over the sensor) or a spike (if the dust scatters insolation into the sensor). In either case, the measured insolation excursion relates directly to the dust devil's optical depth $\tau$ (in the limit of small $\tau$). \citet{LORENZ20151} deployed such an instrument suite on a terrestrial playa and found 20\% of events caused dimming greater than about 2\%. The encounters without detected attenuation may either be dustless vortices or the encounter geometry simply did not produce a signal. Stronger dimming was associated with larger pressure drops, presumably more vigorous and therefore windier vortices. Understanding the relationships between a vortex's pressure and wind profiles and its dust content is critical for accurately estimating the contribution of dust devils to the martian atmospheric dust budget, key to Mars' climate \citep{2004JGRE..10911006B}.

The initial release of data from Mars 2020's MEDA PS (pressure) and RDS (radiation and dust) instrument provide an opportunity to explore these relationships in a novel locale on Mars. Moreover, since vortex formation depends on ambient meteorological conditions \citep{2016SSRv..203..183R}, assessment of their occurrence rate provides a probe of Mars' boundary layer. Fortunately, \citet{2021SSRv..217...20N} recently conducted a comprehensive survey of model predictions for the meteorology within Jezero Crater and observed by Perseverance. That study included predictions of vortex occurrence and suggested that vortices may occur more frequently within Jezero than at other sites hosting recently landed Mars missions, including the InSight mission \citep{Spiga2021} and the Mars Science Laboratory (MSL) Curiosity rover \citep{2021Icar..35914207K}.

In this study, we analyze the MEDA data from the Mars 2020 Perseverance mission released on 2021 Aug 20 to assess the rates of vortex and dust devil occurrence. Over the 89 sols of currently available mission data, we estimate Mars 2020 encountered at least \numvortices\ vortices, one quarter of which induced statistically significant excursions in insolation as observed by RDS. Unfortunately, this same data release did not include fully processed wind data from MEDA WS, meaning uncorrected biases may pervade the data, and so an analysis of winds associated with these encounters must await future studies. We also leave the cache of images collected by Mastcam-Z and Perseverance's engineering cameras for future analysis. Our preliminary assessment, however, provides a catalog of detections for use by subsequent studies in a similar vein to other work \citep[e.g.,][]{2021Icar..35514119L}, and the results comport with meteorological predictions: Jezero Crater seems to be significantly more active than the InSight landing site or Gale Crater, as Mars 2020 encountered nearly 5 vortices per sol on average. 

\section{Data and Model Analysis} \label{section:Data and Model Analysis}

\subsection{Pressure Sensor Data Analysis and Modeling}
\begin{figure}
    \centering
    \includegraphics[width=\textwidth]{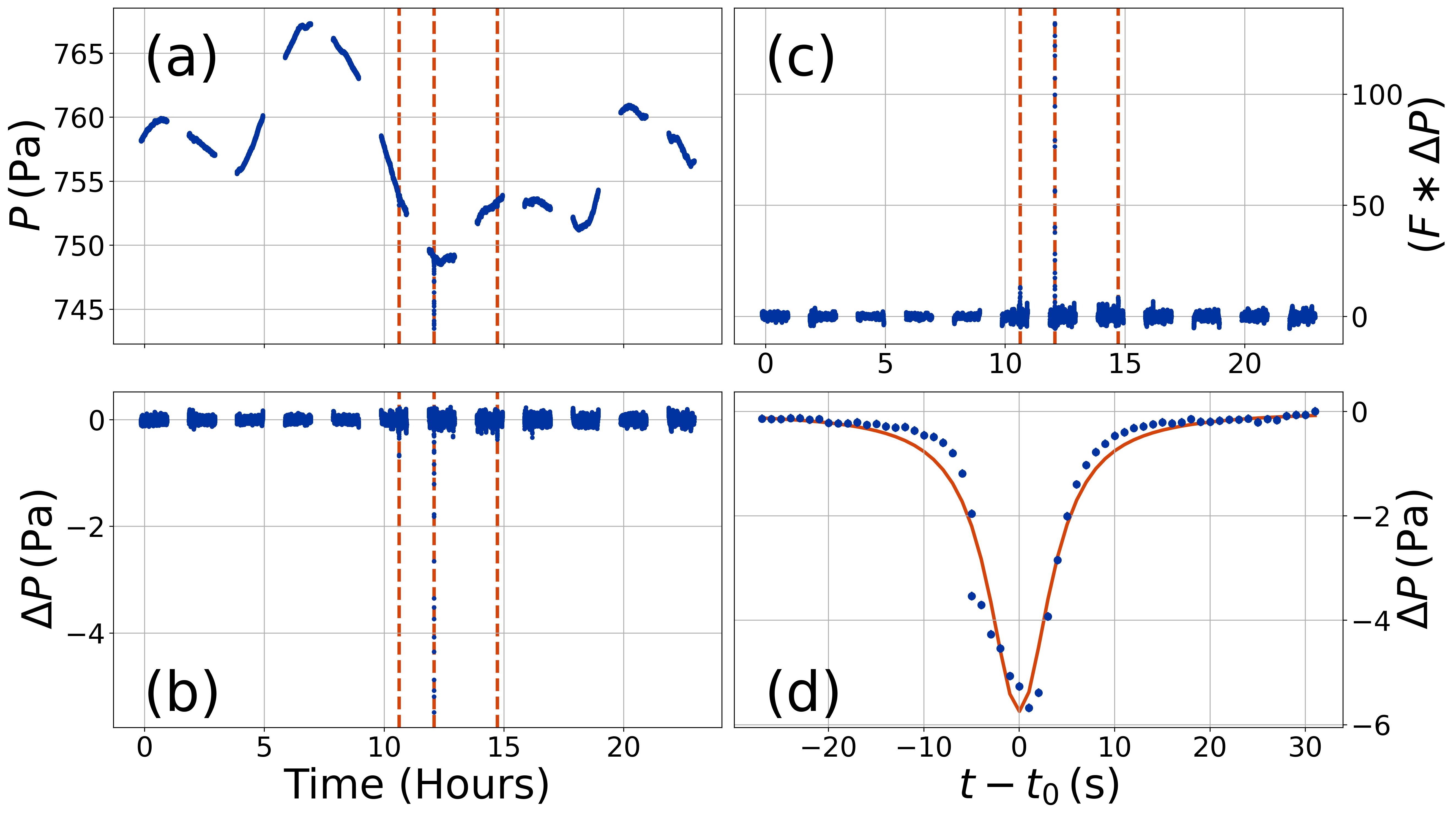}
    \caption{(a) The pressure time-series for sol 82, as blue dots. The vertical orange lines highlight the detected vortex signals. (b) The time-series after application of the mean boxcar filter. Apparent by eye, the scatter $\sigma_{\Delta P}$ in the time-series increases around mid-day. (c) Convolution of the matched filter with the time-series in (b). (d) A model fit (solid orange line) to the deepest vortex discovered on sol 82. Uncertainties are shown but are barely larger than the point diameters.}
    \label{fig:conditioned_data_sol82}
\end{figure}

The analysis presented here follows closely the process employed in \citet{2021arXiv210712456J}. For the present study, we analyzed pressure time-series from the PS instrument available from NASA PDS (\url{https://pds-atmospheres.nmsu.edu/PDS/data/PDS4/Mars2020/mars2020\_meda/}). We used the data set labeled ``data\_derived\_env'' since it represents the most completely processed and calibrated data set (see the Mars 2020 MEDA PDS Archive Bundle Software Interface Specification for details -- \url{https://pds-atmospheres.nmsu.edu/PDS/data/PDS4/Mars2020/mars2020\_meda/document/meda\_bundle\_sis.pdf}). \citet{2021SSRv..217...48R} provides many details on the processing and calibration of these data and relevant references. The data are divided up by mission sol and typically span from midnight one sol to midnight the next, with a sampling rate of $1\, {\rm Hz}$; however, the data for all sols include gaps of at least an hour or more. Some sols (early in the mission) span only a few hours, sols 1, 2, 4, 9, 14, 19, and 22. Sol 10 does not seem to have a pressure time-series at all. Thus, we excluded all these sols from our analysis. Figure \ref{fig:conditioned_data_sol82}(a) shows a representative raw pressure time-series from sol 82 of the mission. 

We do not include the MEDA wind data in our analysis here. The first MEDA data release did not provide the ``derived'' wind data. These ``calibrated'' wind data have not been fully processed to correct for, for example, perturbations on the measured winds from the lander body itself \citep{2021SSRv..217...48R}. However, many prior studies of martian vortices have used only pressure time-series to produce important results, and these studies have employed the assumption that vortex-like pressure dips are vortices. This study follows in that same vein.

We model the pressure signal of a vortex, dustless or dusty, as in \citet{2021arXiv210712456J}:
\begin{equation}
    \Delta P(t) = \frac{-\Delta P_0}{1 + \left( \frac{ t - t_0 }{\Gamma/2} \right)^2}\label{eqn:Lorentzian_profile}
\end{equation}
where $t$ is the time, $t_0$ is the time of closest approach, $\Delta P_0$ is the excursion amplitude, and $\Gamma$ is the observed profile full-width/half-max (FWHM). The PS pressure time-series exhibit a variety of variations on timescales of hours to days, related to meteorological phenomena other than vortices \citep[cf.][]{2020SSRv..216..148P, 2021SSRv..217...20N}. These other signals obscure the short-lived vortex signals, and so we applied a high-pass boxcar filter to the raw pressure data with a window size of \boxcarwindowsize\ to suppress the long-term variability. Experimentation with the data showed this window size provided a reasonable balance between flattening the long-term signals without significantly distorting the vortex signals. As for all data-processing schemes \citep{2018Icar..299..166J}, this one will inevitably suppress or distort some signals, including vortex signals spanning longer than \boxcarwindowsize. The typical martian vortex is tens of meters in diameter with an advection speed of a few $\mathrm{m\ s^{-1}}$ \citep{2016SSRv..203..277L}, so we do not expect a \boxcarwindowsize\ cutoff to significantly skew our inferred population. We also conduct injection-recovery experiments to assess these biases in Section \ref{section:Results} (see Figure \ref{fig:injection-recovery_sol82}). In any case, the resulting detrended pressure time-series exhibit standard deviations $\sigma_{\Delta P}$ between $0.05$ and $0.08\,{\rm Pa}$. Figure \ref{fig:conditioned_data_sol82}(b) shows the detrended dataset for sol 82. 

To recover the vortex signals, we then applied a matched filter with a shape given by Equation \ref{eqn:Lorentzian_profile} to the detrended data. In other words, we marched a Lorentzian profile, point-by-point, across the time series, convolving it with the time-series, producing a convolution signal $F\ast\Delta P$. We then subtracted the mean value of the resulting convolution signal and divided by the standard deviation to scale the spectrum by the intrinsic noise in the dataset. (The noise arises from a combination of turbulent pressure fluctuations not associated with vortex detections, instrumental effects, and sampling rate, and using the standard deviation as an estimate for such noise is a traditional, albeit limited, approach -- \citealp[cf.][]{2003drea.book.....B}.) When a short-lived, negative pressure excursion occurs in the time-series, the convolution signal shows large positive spike. Experimentation with the data suggested a threshold value of $F\ast\Delta P \ge$ \detectionthreshold\ provides a good balance between excluding spurious or doubtful excursions and recovering statistically significant excursions. Figure \ref{fig:conditioned_data_sol82}(c) shows the convolution signal for the data in panel (b). Vertical, dashed orange lines show the spikes exceeding our detection threshold.

Finally, using the Levenberg-Marquardt algorithm \citep[cf.][]{Press2007}, we fit each putative vortex signal from each sol's data to retrieve best-fit $t_0$, $\Delta P_0$, and $\Gamma$-values. To avoid signal distortion from our detrending process, we fit the original, un-detrended data (Figure \ref{fig:conditioned_data_sol82}a), which required us also to add a background linear trend to the vortex signal itself. We used the standard deviation for each sol's detrended pressure time-series as the per-point uncertainties. Uncertainties on model parameters are given by the square root of the variable covariance matrix, scaled by the square root of the reduced $\chi^2$-value for the model fit, effectively imposing $\chi^2 = 1$ \citep{Press2007}. As an example of the clearest signal, Figure \ref{fig:conditioned_data_sol82}(d) shows the deepest vortex we detected, with $\Delta P_0 = \left( 5.7 \pm 0.9 \right)\,{\rm Pa}$ and $\Gamma = \left( 7 \pm 2 \right) \,{\rm s}$.

\subsection{Radiation and Dust Sensor Data Analysis and Modeling}
We also conducted a preliminary analysis of the RDS radiometric time-series. As described in \citet{2021SSRv..217...48R}, the RDS suite includes 16 independent sensors designed to measure (or to help calibrate the measurements of) the upward and downward-welling radiation. The discrete photodetectors, numbered 1 through 8, are arrayed in a circle on the RDS assembly, some on top and pointed up at the sky (``TOP'') and others pointed outward from the assembly laterally (``LAT''). The eight LAT sensors all sample a narrow band of wavelengths $750\pm10\,{\rm nm}$, while the TOP sensors each sample different bands, spanning from $190\,{\rm nm}$ to $1100\,{\rm nm}$. Sensor LAT\_1 is blocked by the Sampling and Caching Subsystem (SCS), and so the mission has blinded it to assess the degradation of all the sensors due to radiation \citep{2021SSRv..217...48R}. The detectors (except LAT\_1) are sensitive to scattering of light by the martian dust. In the current data release, the RDS dataset has not been completely processed (i.e., the data are categorized as ``calibrated'' -- \url{https://pds-atmospheres.nmsu.edu/PDS/data/PDS4/Mars2020/mars2020\_meda/document/meda\_bundle\_sis.pdf}). 

\begin{figure}
    \centering
    \includegraphics[width=\textwidth]{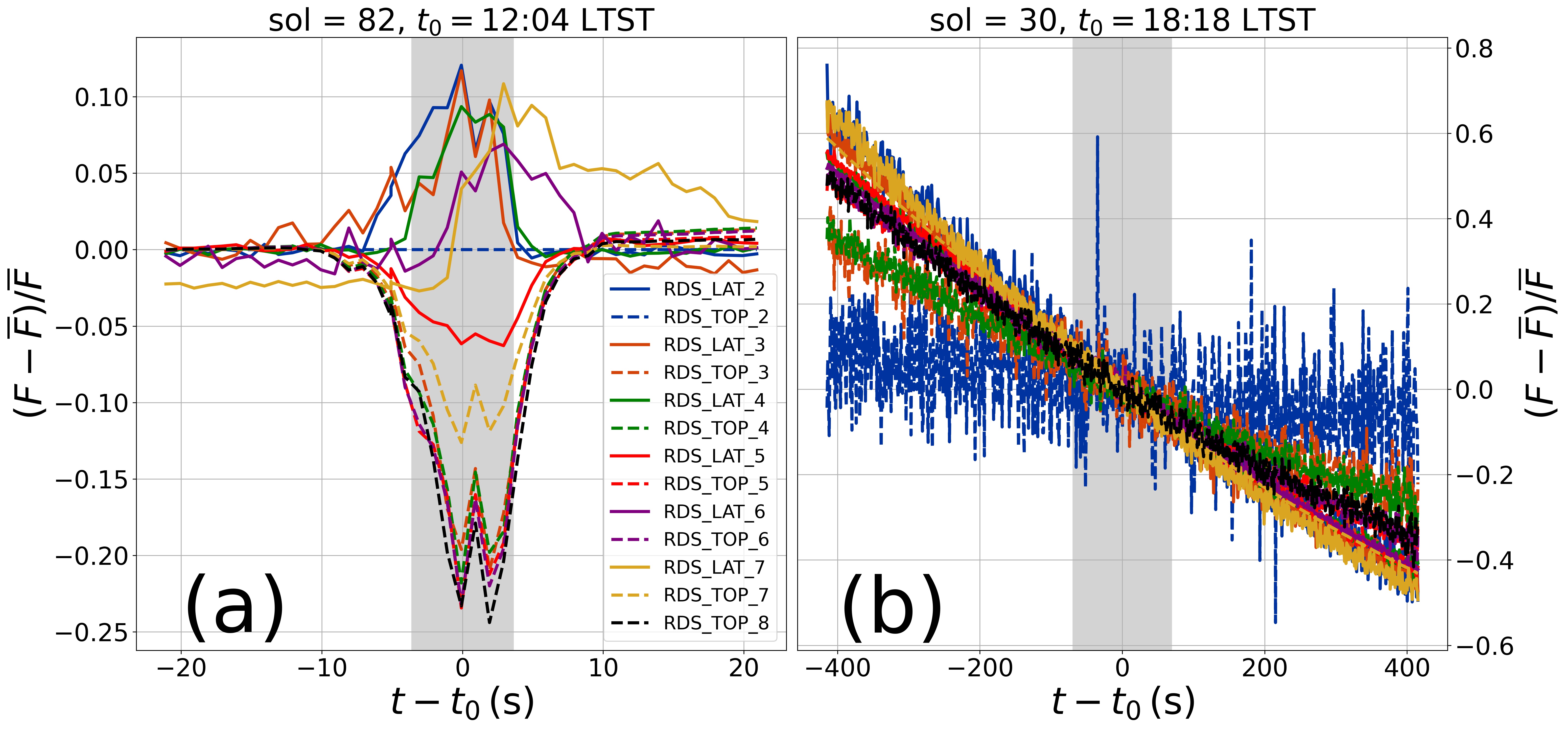}
    \caption{(a) RDS time-series collected during the vortex encounter on sol 82 at $t_0 =$ 12:04 LTST. Each combination of line color and styles reflects a specific sensor as indicated in the legend. The filled grey band shows the FWHM ($\Gamma$) for the vortex encounter. (b) The same as in (a), except for an encounter on sol 30 at $t_0 =$ 18:18 LTST, showing no statistically significant excursion. }
    \label{fig:vortex_RDS_example}
\end{figure}

The passage of a dust-laden vortex near the RDS sensors registers in the time-series as a series of (negative) dips and (positive) blips, as the dust scatters light into and away from the sensors \citep{LORENZ20151}. The structures of these signals may be complicated, reflecting the potentially complex column density structure within the vortices -- Figure \ref{fig:vortex_RDS_example} provides examples during two vortex encounters, showing data from all the RDS sensors (except LAT\_1). 

In the optically thin limit, the magnitude of the excursion (positive or negative) scales with column optical depth, so we consider the maximum excursion during an encounter from among all the sensors as measured relative to the signal 3 $\Gamma$ before and after the time of encounter $t_0$. We estimate the median value of the RDS signals $\overline{F}$ and the scatter $\sigma_{F}$ from these before and after periods. If the maximum of the absolute value of signal during the encounter exceeds 3$ \sigma_{F}$, we recorded this value ${\rm max} | \frac{F - \overline{F}}{\overline{F}} |$ as a statistically significant excursion. Other encounters were assigned a value of zero. (Because some of the resulting excursions appeared spurious, we filtered out values more than five standard deviations larger than the median value for all excursions.) No statistically significant excursion may represent either a dustless vortex or an encounter for which the light-scattering geometry simply did not produce an excursion (e.g., no shadow fell across the sensors).

Since we are interested here only in looking for excursions in the RDS signals rather than in a detailed modeling effort, the ``calibrated'' dataset (rather than the ``derived'' dataset) suffices for our purposes -- see \url{https://pds-atmospheres.nmsu.edu/PDS/data/PDS4/Mars2020/mars2020\_meda/} for more details. Even this preliminary analysis, however simplistic, is a significant step forward compared to prior vortex analyses: most prior landers did not even have instruments that could register variations in insolation corresponding to dust devil passages, and those that did produced datasets with severe limitations \citep{2016Icar..278..180S}. Analyses following ours can and should improve on our approach, but the results here represent a key first step, providing a catalog of initial detections upon which these subsequent studies may be based.

\section{Results} \label{section:Results}

\begin{figure}
    \centering
    \includegraphics[width=\textwidth]{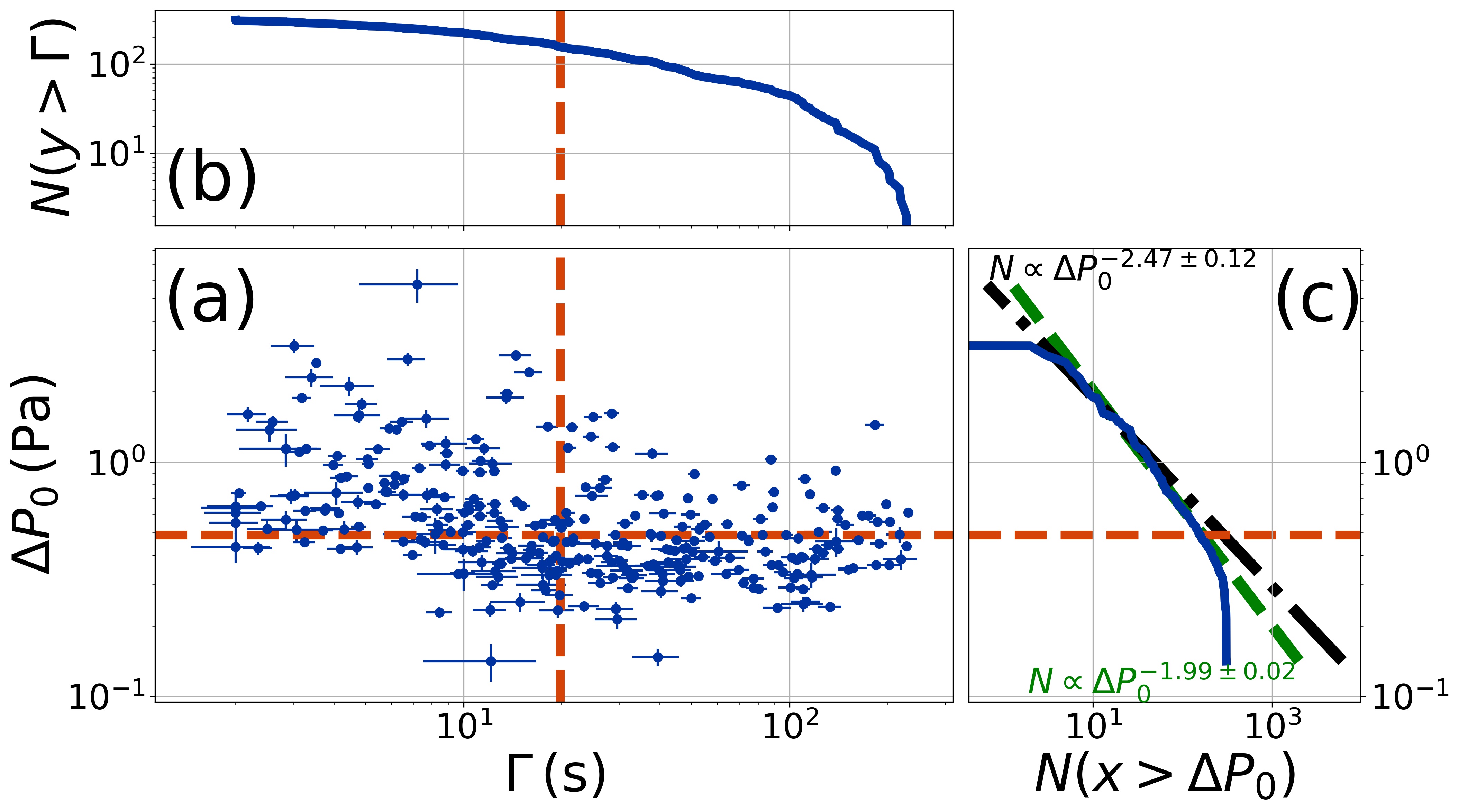}
    \caption{(a) The best-fit $\Delta P_0$- and $\Gamma$-values (blue dots) with error bars. (b) Cumulative histogram of $\Gamma$-values, along with the median value ($\Gamma = \left( 20 \pm 2 \right)\,{\rm s}$) shown by the dashed, orange line. (c) Cumulative histogram of $\Delta P_0$-values, along with the median value ($\Delta P_0 = \left( 0.49 \pm 0.02 \right)\,{\rm Pa}$) shown by the dashed, orange line. The dashed green line shows a power-law fit to the histogram for $\Delta P_0 > 0.5\, {\rm Pa}$ with $N \propto \Delta P_0^{-1.99\pm0.02}$, while the dash-dotted black line shows a fit for $P_0 > 1.5\, {\rm Pa}$ with $N \propto \Delta P_0^{-2.47\pm0.12}$.}
    \label{fig:DeltaP-Gamma_scatterplot-histogram}
\end{figure}

Figure \ref{fig:DeltaP-Gamma_scatterplot-histogram} illustrates the resulting best-fit $\Delta P_0$- and $\Gamma$-values for the collection of \numvortices\ recovered vortex signals that we retained throughout this study, and Table \ref{tbl:vortex_fit_params} provides a complete list of the values. (After applying the matched filter and model-fitting analysis described above, we discarded \discardedvortices\ putative vortex signals with apparent $\Gamma > 250\,{\rm s}$ and $\Delta P_0/\sigma_{\Delta P} < 5$, which consistently seemed to be either spurious detections or spikes resulting from the edge effects from our detrending process.)

As in previous studies \citep[e.g.,][]{2021Icar..35514119L}, we fit power-laws to the cumulative histogram of $\Delta P_0$-values. As discussed next, assessing our ability to recover synthetic vortices with known $\Delta P_0$- and $\Gamma$-values indicates we could consistently recover vortices with $\Delta P_0 \gtrsim 0.6\,{\rm Pa}$ (\vorticesabovelowercutoff\ vortices), and so we used the Levenberg-Marquardt algorithm to fit the power-laws using only those deeper vortices. Poisson sampling was assumed to estimate bin uncertainties. The cumulative histogram shows an apparent knee at about $\Delta P_0 = 1.5\,{\rm Pa}$, so we also fit a power-law to these vortices, although this result is likely skewed by small number statistics (we found only \vorticesaboveuppercutoff\ such vortices). However, a similar knee was reported in \citet{Spiga2021} for vortices encountered by InSight.

\begin{figure}
    \centering
    \includegraphics[width=\textwidth]{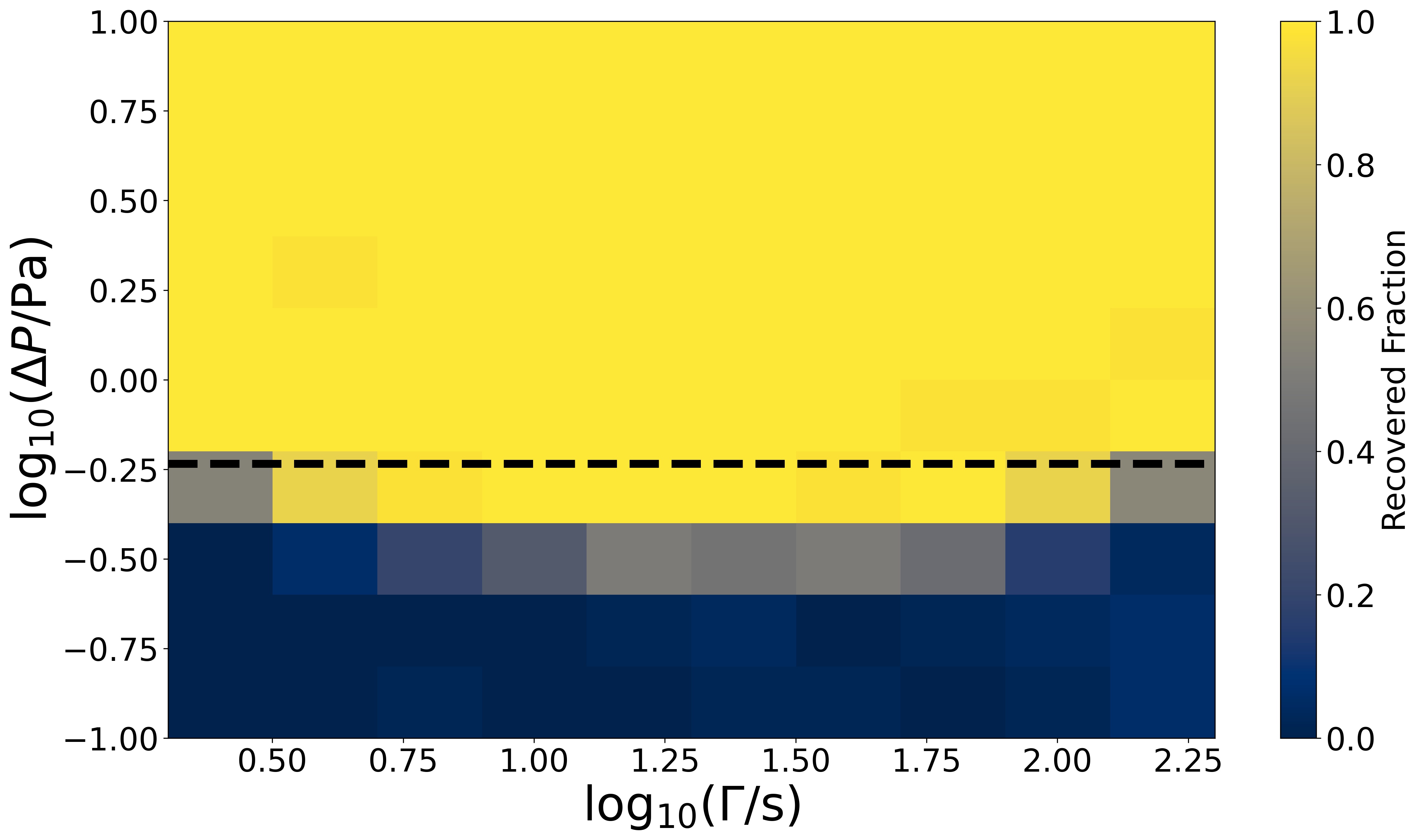}
    \caption{Results from an injection-recovery calculation using data from the sol with the largest scatter in $\Delta P$ ($0.08\,{\rm Pa}$), sol 82. The dashed black line shows the detection threshold, which corresponds to $\Delta P_0 = 0.6\,{\rm Pa}$. All but the very shortest or longest duration vortices with $\Delta P_0$ above that threshold are consistently recovered.}
    \label{fig:injection-recovery_sol82}
\end{figure}

In order to assess the robustness of our detection scheme, we conducted several injection-recovery experiments for which we injected vortex signals with known parameters into the raw pressure time-series and then applied our scheme to determine how often we could successfully recover the vortices. Figure \ref{fig:injection-recovery_sol82} shows the results for the sol with a time-series exhibiting the largest scatter, sol 82. We find that we can consistently recover signals with $\Delta P_0 = 0.6\,{\rm Pa}$ but that the scatter often obscures less deep signals. Thus in our power-law analysis, we focused on the deeper vortex signals.

Our results here are roughly consistent with recent analyses of data from the InSight mission. Analyzing data from the mission's pressure sensor APSS, \citet{Spiga2021} reported a power-law index for vortex detections with $0.35\,{\rm Pa} \le \Delta P_0 \le 9\,{\rm Pa}$ consistent with $-2.4\pm0.3$ for the cumulative histogram. \citet{2021Icar..35514119L} conducted an analysis of the same dataset and, considering vortices with $0.8\,{\rm Pa} < \Delta P_0 < 3\,{\rm Pa}$, found a power-law index of $-2$. For deeper vortices, an index of $-3$ was suggested to provide a better fit. \citet{2021arXiv210712456J} reported an overall index of $-2.39\pm0.02$ for vortices with $\Delta P_0 > 1\,{\rm Pa}$.

\begin{figure}
    \centering
    \includegraphics[width=\textwidth]{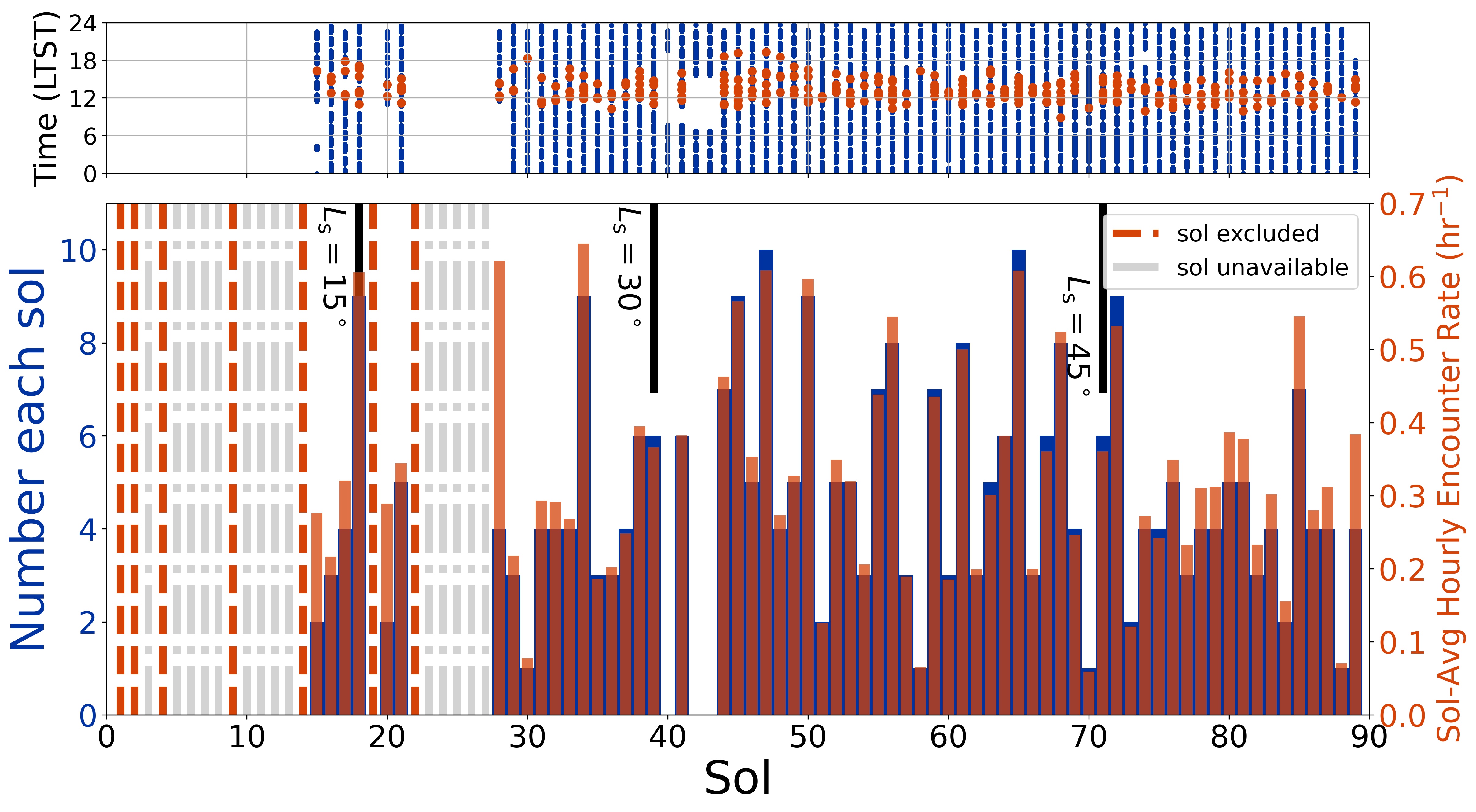}
    \caption{(Top) When pressure data were collected during each sol. The orange dots also show when vortex signals were detected. (Bottom) The blue bars show the number of vortex encounters during each sol, while the orange bars show the number of encounters each sol divided by the total number of hours during which the pressure sensor collected data. The dashed orange lines show sols when pressure data were available but were flawed or inadequate and so excluded for our analysis. The grey dash-dotted lines show sols when data were unavailable. (For the sols near 40 with no bars, we found no encounters, probably because no data were collected near mid-day on those sols.) Solid black lines indicate $L_{\rm s}$-values.}
    \label{fig:vortex_occurrence_each_sol}
\end{figure}

\begin{figure}
    \centering
    \includegraphics[width=\textwidth]{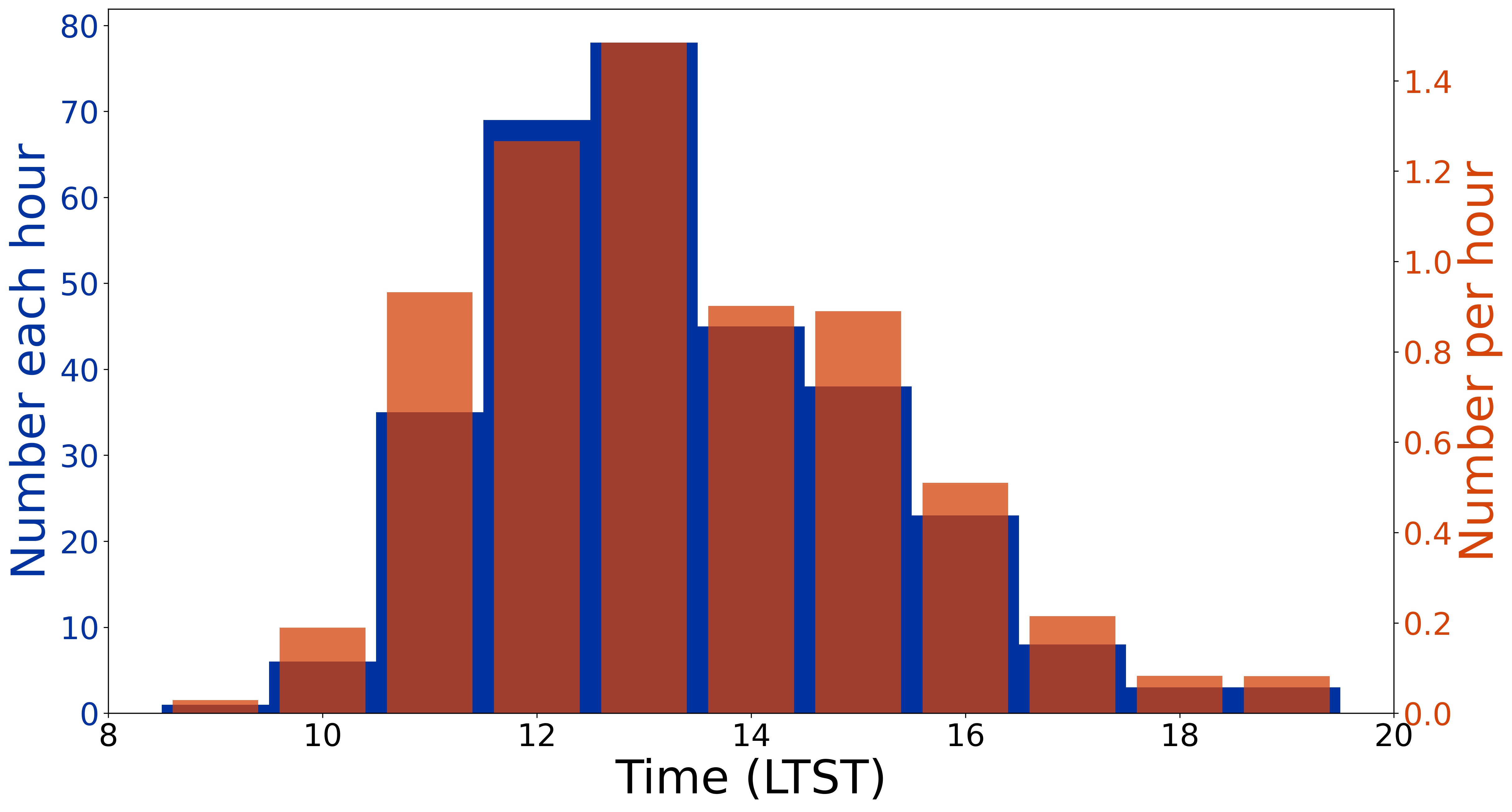}
    \caption{The blue bars show the total number of vortex encounters that took place during that hour over the whole 89 sol dataset, while the orange bars show that number divided by the total number of hours during that period each sol observed throughout the 89 sol dataset.}
    \label{fig:vortex_occurrence_each_hour}
\end{figure}

We can also explore the vortex encounter rate for Perseverance. Figure \ref{fig:vortex_occurrence_each_sol} show the encounters binned by sol as blue bars. As indicated above, several sols have no data available, and others have data problematic for our study, as indicated in the figure. For the 65 sols with available, usable data, there was an average rate of $5\pm2$ encounters per sol (where error bars come from assuming Poisson statistics), with variations between 0 and 10. Some sols had more data available than others, however (some sols had as few as 7 hours of observations, while others had as many as 17 hours). To account for that variation, we divided the number of encounters on a given sol by the total number of hours (or fractions thereof) of observational data, giving the orange bars in Figure \ref{fig:vortex_occurrence_each_sol}. For example, Figure \ref{fig:vortex_occurrence_each_sol} shows there were 9 vortex encounters on sol 50. There were about 15 hours total of pressure logger data available for that sol, which implies an encounter rate of about 0.6 per hour (9/15) on that sol. (Some previous studies calculate an encounter rate by normalizing only by the number of hours during a sol when vortices were observed, not by all the hours during a sol. However, that approach involves some uncertainty since it depends on the observationally defined time-of-day when the latest vortex is detected. Instead, we normalized by the total number of hours, which reduces somewhat the sol-averaged hourly encounter rate compared to other studies.) Some of the sol-to-sol variation is clearly due to variability of when observations were made. The lack of vortex detections around sol 40, for instance, very likely results from the dearth of data collection around mid-day, when vortices are most active. We can see that the typical sol saw an average of one encounter about every three hours, 0.4 encounters per hour (at least while data were collected) up to a maximum of one every 90 minutes (about 0.6 encounters per hour).

The encounter rates also show hour-to-hour variation, as illustrated by the blue bars in Figure \ref{fig:vortex_occurrence_each_hour}. We have normalized the number of encounters during each hour by the total number of hours (over all available/usable sols) to estimate the hourly encounter rate. As seen in previous studies \citep[e.g.,][]{2021arXiv210712456J}, the encounter rate peaks about mid-day, in this case at 1.5$\pm 0.2$ encounters per hour (once every 40 minutes), dropping below detectable levels early in the morning and late in the afternoon; however, encounters persist until 19:00 LTST. 

\begin{figure}
    \centering
    \includegraphics[width=\textwidth]{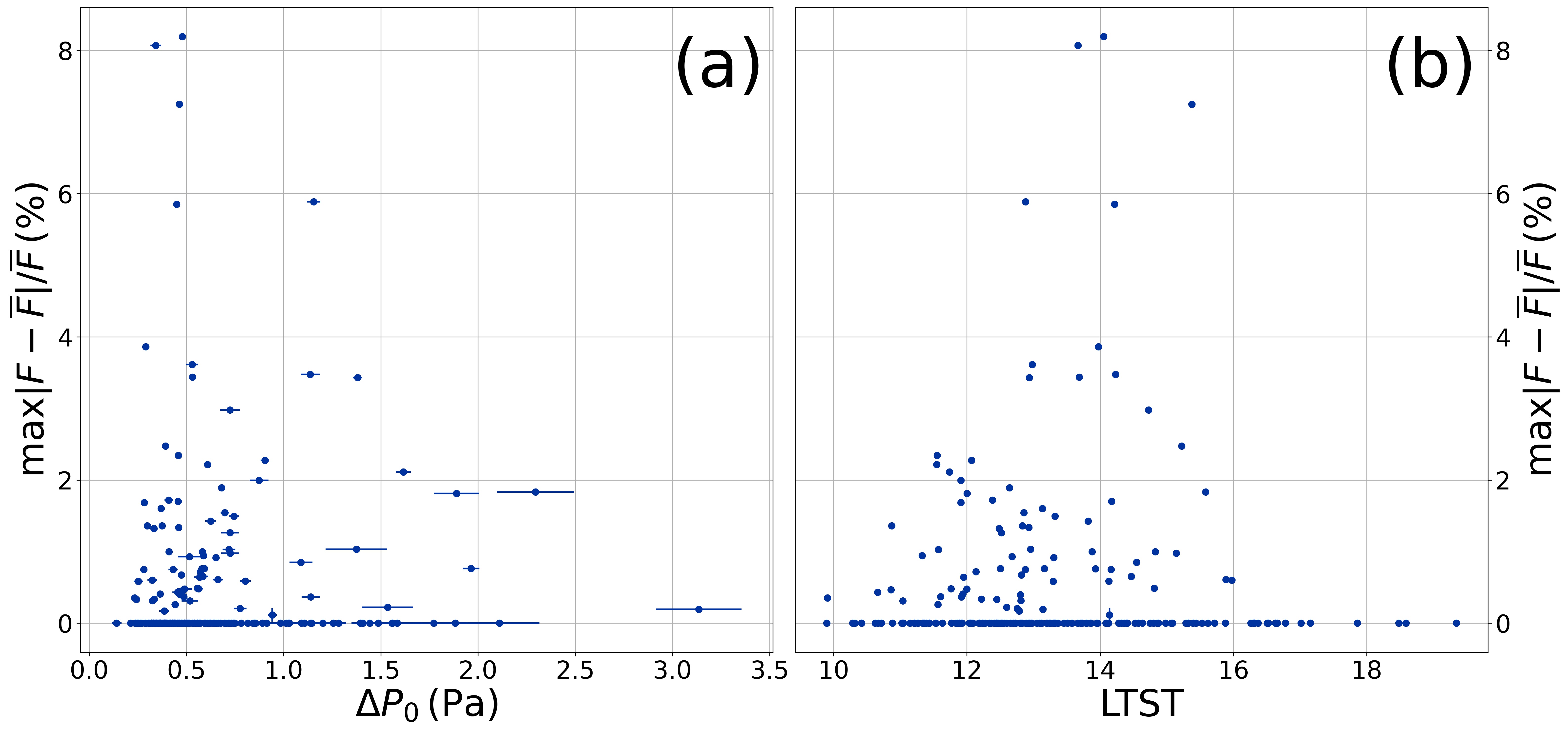}
    \caption{(a) RDS excursions vs. the observed $\Delta P_0$-value for the vortex encounters. (b) RDS excursions vs. the time-of-sol for the encounter. Error bars for all estimated variables are shown but are smaller than the plot symbol in most cases.}
    \label{fig:max_RDS_excursion}
\end{figure}

Small-number statistics may influence the results reported here. \citet{2015JGRE..120..401J} discussed the influence of small-number statistics on the inferred population of vortices, finding, for example, that the best-fit power-law for the pressure histogram may depend sensitively on the population size, with some bias toward inaccurately inferring a more shallow distribution (i.e., the derived power-law index was closer to zero than it would be if a larger population were recovered). Additional vortex detections by Mars 2020 will undoubtedly improve upon the results presented here, but the agreement with prior results lends credibility to the current results.

\subsection{Are vortices at Jezero Crater more active than at the landing sites of other recent missions?}
At least for the seasons observed so far, these results corroborate predictions that vortex activity at Jezero would exceed activity at Gale Crater, the exploration site for the Mars Science Laboratory rover (MSL) Curiosity, and likely at the landing site for the InSight mission. The comprehensive results in \citet{2021SSRv..217...20N} predicted significantly elevated vortex activity based on the high values estimated for the ``dust devil activity'' DDA. This parameter combines a measure of the thermodynamic efficiency of dust devil convection with the near-surface sensible heat flux, two parameters thought to be key for vortex activity \citep{1998JAtS...55.3244R}. How exactly this parameter maps to vortex occurrence (Does larger DDA mean more vortices overall? More vigorous vortices?~etc.) remains unclear, but DDA is a metric derivable from general circulation models and large eddy simulations and likely has some direct relationship to vortex occurrence. Thus, a fuller understanding of its relationship to vortex properties would elucidate key boundary layer processes.

\citet{2019JGRE..124.3442N} analyzed three Mars years of pressure and wind time-series from Curiosity, spanning sols 1 to 1980 of the mission, and estimated vortex encounter rates that varied from sol to sol, season to season. Considering pressure excursions greater than $0.6\,{\rm Pa}$, per-sol encounter rates topped out at about 4 per sol (during the summer), with a typical non-zero value of 1 per sol. The blue bars in Figure \ref{fig:vortex_occurrence_each_sol} show that the typical number of vortices encountered each sol was comparable to the maximum number for Curiosity. That study also reported per-hour encounter rates that topped out at about 1 per hour but was more typically 0.5 per hour when it was not zero, both with uncertainties of about 0.1. Figure \ref{fig:vortex_occurrence_each_hour} shows a maximum number per hour about 40\% larger than the largest encounter rate for Curiosity, with an average (non-zero) value of about 0.6$\pm 0.2$ per hour, 20\% larger than the average value for Curiosity. \citet{2018Icar..299..308O} conducted a similar survey of the first two Mars years of pressure data from Curiosity and found similar encounter rates for pressure excursions exceeding $0.5\,{\rm Pa}$. Comparing to the results in Figure \ref{fig:vortex_occurrence_each_sol} and \ref{fig:vortex_occurrence_each_hour}, it seems clear that Perseverance saw more vortices with similar minimum pressure excursions. 

With regard to the InSight results, there appears to be some mismatch between different studies. \citet{Spiga2021} analyzed the first 400 sols of InSight pressure time-series data and reported more than 6000 vortex encounters with pressure excursions exceeding $0.35\,{\rm Pa}$. Considering only encounters with an excursion exceeding $0.5\,{\rm Pa}$, they reported an overall encounter rate of 7 per sol, with (non-zero) rates varying between 1 and 17 per sol. \citet{2021Icar..35514119L} also sifted the InSight pressure and wind data for vortex encounters. Considering a minimum excursion of $0.8\,{\rm Pa}$, that study reported an overall rate of 2-3 encounters per sol, which, using the results of the histogram analyses from that study (their Figure 4), works out to about 5 encounters per sol for an excursion greater than $0.5\,{\rm Pa}$. \citet{2021arXiv210712456J} conducted an independent analysis of a slightly enlarged InSight dataset (477 sols worth) and found encounter rates topping out at 3 per sol and 0.4 per hour for excursions exceeding $0.3\,{\rm Pa}$. \citet{2021arXiv210712456J} discussed possible reasons for the mismatches between these studies. In any case, the per-sol encounter rates reported here for Perseverance (Figures \ref{fig:vortex_occurrence_each_sol} and \ref{fig:vortex_occurrence_each_hour}) often match or exceed all but the highest rates reported in those previous studies. (\citealt{Spiga2021} and \citealt{2021Icar..35514119L} do not report per-hour encounter rates.) As the Mars 2020 mission continues, more vortex detections will stack up, providing more robust estimates. It is possible that seasonal variability will bring the encounter rate at Mars 2020 down to or below the rates seen at other sites, although models suggest the encounter rate will increase \citep{2021SSRv..217...20N}.


The fractional area covered by dust devils has been cited as a useful metric for estimating the likelihood of vortex encounters for a landed spacecraft. \citet{2021Icar..35514119L} pointed out that the fractional area can be estimated by comparing the total durations of all vortex encounters to the total observational time. The total duration of encounters reported here is 3.56 hours, while the total observational time over the 89 sols is about 966 hours. Together, these numbers suggest a fractional area of 0.367\%, about five times larger than the fractional area seen for vortices at InSight \citep{2021Icar..35514119L, 2021arXiv210712456J}.

\subsection{What fraction of encountered vortices are true dust devils?}
Figure \ref{fig:max_RDS_excursion} provides some insight into how often vortices are dusty. As previously indicated, about one quarter (\numRDSexcursions\ of our total \numvortices) produced a discernible radiative signal. At face value, this result suggests about one quarter of the encountered vortices lofted dust at the level it could be detected. This fraction compares favorably to results from \citet{LORENZ20151}. That study involved a terrestrial deployment of pressure loggers and solar insolation sensors and found that about 20\% of vortex encounters exhibited insolation excursions of 2\% or greater (a 30\% excursion in one case). Of course, the actual geometry of the encounter dictates whether a dust vortex will produce an excursion (Does the vortex pass on the sunward or anti-sunward side of the sensor?), but there is no obvious reason for the encounter geometries from \citet{LORENZ20151} to differ substantially from the geometries for Perseverance, at least not at first order.

The fraction of apparently dusty vortices reported here also closely matches the maximum fraction inferred in \citet{2021arXiv210712456J} for the InSight Mission. As in \citet{Spiga2021}, that study reported no visual detections of dust devils, and the InSight lander does not include any insolation sensors. However, \citet{2021arXiv210712456J} used the lack of imaged dust devils, convolved with the frequency of imaging and of vortex encounters, to infer an upper limit for the fraction of vortices lofting measurable amounts of dust at 35\%. 

The conditions that allow a vortex to loft dust remain obscure, in spite of decades of field studies, laboratory experiments, and modeling \citep{2016SSRv..203..183R}. However, the results here can help shed some light. Among vortices with measurable radiative excursions in Figure \ref{fig:max_RDS_excursion}, the encounter with the smallest excursion that is also more than 5$ \sigma_F$ greater than zero (meaning the excursion is credibly non-zero) is ${\rm max} |F - \overline{F}|/\overline{F} = 0.0017\pm5\times10^{-6}$ and has $\Delta P_0 = 0.39\,{\rm Pa}$. Assuming cyclostrophic balance \citep{2016SSRv..203..209K}, this pressure deficit corresponds to an eyewall velocity of about $4\,{\rm m\ s^{-1}} (= \sqrt{0.39\,{\rm Pa}/0.02\,{\rm kg\ m^{-2}}})$, well below the expected threshold for dust-lifting on Mars, between 20 and $30\,{\rm m\ s^{-1}}$ \citep{2003JGRE..108.5041G}. (With the right geometry, however, an ambient wind could contribute to the vortex's lifting power.) An eyewall velocity of $20\,{\rm m\ s^{-1}}$ corresponds to a central pressure dip for a vortex of $\Delta P = 8\,{\rm Pa} (= (20 {\rm m\ s^{-1})}^2 \times (0.02\,{\rm kg\ m^{-3}}))$. We can see from Figure \ref{fig:max_RDS_excursion} that none of our encounters registered such a large pressure excursion. These results suggest that, if $20\,{\rm m\ s^{-1}}$ is the true lifting threshold, all of our apparently dusty vortices were encountered off-center, giving observed pressure minimum $\Delta P_0$ well below the central values, as expected statistically \citep{2018Icar..299..166J, 2019Icar..317..209K}. 

The dusty vortex with the deepest pressure signal ($\Delta P_0 = 3.1\,{\rm Pa}$) also has one of the smallest radiative excursions. Indeed, there is a dearth of vortex encounters with large $\Delta P_0$ and large radiative excursions in Figure \ref{fig:max_RDS_excursion}(a), which seems counter-intuitive: we might expect the most vigorous vortices to lift the most dust. Theoretical expectations \citep{2020Icar..33813523J} and observations of martian dust devils \citep{2006JGRE..11112S09G} corroborate this expectation. Instead, this dearth may arise simply from the vortex encounter geometries. A relatively low pressure (small $\Delta P_0$) signal may result either from a (more likely) distant encounter with a vigorous (large central pressure deficit) and very dusty vortex \emph{or} from a nearby encounter with a weak (small central pressure deficit) and low-dust vortex. The former encounter may result in a large RDS excursion, the latter in a small excursion. The spread in RDS excursions for the smallest $\Delta P_0$-values in Figure \ref{fig:max_RDS_excursion} appears to corroborate that expectation. By contrast, a relatively high pressure signal (large $\Delta P_0$) is most likely to result from a nearly central encounter. During such an encounter, the Sun will only be occulted by one wall of the dust devil, resulting in a relatively low optical depth encounter. In any case, the completely processed wind data from MEDA (when available) will be crucial for understanding these trends since the wind speed and/or direction measurements may allow independent determination of the encounter geometry \citep{2016Icar..271..326L, 2021Icar..35914207K, 2021arXiv210712456J}.

Figure \ref{fig:max_RDS_excursion}(b) reveals a pattern easier to interpret. The largest RDS excursions occur at very nearly the same time of day as the peak in occurrence rate. Since DDA also peaks near 13:00 LTST \citep{2021SSRv..217...20N}, these results seem to suggest that larger values of DDA correspond both to increased vortex encounter rates and enhanced dust lofting. DDA increases, in part, with as the boundary layer deepens, which likely results in taller dust devils. \citet{2020Icar..33813523J} suggested that taller dust devils ought to have larger dust densities, and the results here comport with that prediction.

\section{Conclusions}
\label{sec:Conclusions}

This study presents a preliminary analysis of vortex and dust devil encounters from the first 89 sols of data from the Perseverance rover's MEDA meteorological suite. Although some key data are not yet completely processed or available, including the wind measurements, we can draw some tentative but intriguing conclusions. The distribution of observed pressure excursions for the vortex encounters satisfies a power-law fit in agreement with other analyses \citep[cf.][]{2016SSRv..203..277L}. The hour-by-hour encounter rate varies throughout the sol with a peak near mid-day, again, similar to previous observational studies \citep{2016SSRv..203...39M} and model predictions tailored to Jezero Crater \citep{2021SSRv..217...20N}. 

Our results suggest vortex encounters for Perseverance exceed the encounter rates at Curiosity by between a factor of 5 to 10. The rates likely exceed those at InSight, but there is some disagreement about InSight's precise encounter rate between previous studies. Perseverance's RDS instrument allows us to assess whether a vortex was actually dust-laden as dusty vortices induced positive and negative excursions in the RDS time-series as a result of light-scattering. One quarter of the vortices show signs of dust-lofting.


As additional data are made available and processed from Perseverance, additional insights can be gleaned. Auspiciously, \citet{2021SSRv..217...20N} actually predict higher DDA-values for the summer season, so we might expect even higher vortex encounter rates than reported here, although time will tell whether Jezero Crater is, indeed, a more active site than others. A complete model accounting for temporally and spatially evolving phase angles, complex light-scattering properties of the dust, etc.~\citep[e.g.,][]{2013Icar..223....1M} could return a robust assessment of dust profiles from the RDS time-series. Estimates of the durations of the RDS signals may provide dust devil diameters. Analysis of the wind speed and directional data can also elucidate vortex diameters and even encounter geometries. A detailed survey of imagery from the engineering cameras and Mastcam-Z can constrain dust devil frequency, diameters, and dustiness \citep{2010JGRE..115.0F02G}.

\appendix
\section{Vortex Fit Parameters}
Table \ref{tbl:vortex_fit_params} provides the vortex fit parameters. The leftmost column indicates the mission sol on which the vortex encounter took place. The $t_0$ encounter time is given in LTST with uncertainties (in seconds or fractions of a second) shown. The $\Delta P_0$ column shows the maximum estimated vortex pressure excursion, the $\Gamma$ column shows the full-width/half-max, and the rightmost column shows the RDS excursion, when one was detected. In some cases, uncertainties on the fit parameters may be slightly more precise than the instrument precisions because the shape of a vortex signal more precisely constrain the parameters. For example, the second vortex encountered on Sol 15 has a $t_0$ uncertainty of $0.4\,{\rm s}$ even though the pressure sensor only samples at 1-Hz. In general, uncertainties for parameters from a linear fit derived from a dataset depend on the covariance matrix, not just on the data resolution \citep{2003astro.ph.10577G}.



\startlongtable
\begin{deluxetable}{cccccc}




\tablecaption{Vortex fit parameters.}

\tablenum{1}

\tablehead{\colhead{Sol} & \colhead{$t_0$} &
\colhead{$\Delta P_0$} & \colhead{$\Gamma$} & \colhead{${\rm max}|F -
\overline{F}|/\overline{F}$} \\ 
\colhead{} & \colhead{(LTST)} & \colhead{(Pa)} & \colhead{(seconds)} & \colhead{} } 

\startdata
15 &  16:18:50$\pm$3 &  0.42$\pm$0.04 &  61$\pm$14 &  - \\
15 &  16:22:14.5$\pm$0.4 &  0.98$\pm$0.07 &  12$\pm$2 &  - \\
16 &  12:47:08.1$\pm$0.7 &  0.39$\pm$0.03 &  23$\pm$3 &  0.001691$\pm$5e-06 \\
16 &  14:38:12.1$\pm$0.5 &  0.38$\pm$0.01 &  30$\pm$2 &  - \\
16 &  15:23:00.2$\pm$0.2 &  0.64$\pm$0.09 &  2.0$\pm$0.4 &  - \\
17 &  12:16:24.9$\pm$1.0 &  0.57$\pm$0.01 &  81$\pm$4 &  - \\
17 &  12:32:08.1$\pm$0.7 &  0.49$\pm$0.09 &  8$\pm$3 &  - \\
17 &  16:17:54.9$\pm$0.3 &  0.7$\pm$0.01 &  49$\pm$2 &  - \\
17 &  17:51:29.1$\pm$0.1 &  0.368$\pm$0.008 &  13.0$\pm$0.5 &  - \\
18 &  11:00:13.3$\pm$0.3 &  0.42$\pm$0.03 &  10$\pm$1 &  - \\
18 &  12:40:46.5$\pm$0.2 &  0.52$\pm$0.06 &  3.1$\pm$0.6 &  0.009285$\pm$8e-06 \\
18 &  12:49:44$\pm$1 &  0.39$\pm$0.02 &  48$\pm$5 &  - \\
18 &  12:58:38$\pm$1 &  0.24$\pm$0.02 &  29$\pm$4 &  - \\
18 &  13:03:06.4$\pm$0.3 &  0.39$\pm$0.02 &  14$\pm$1 &  - \\
18 &  15:25:14.8$\pm$0.3 &  0.72$\pm$0.01 &  39$\pm$2 &  - \\
18 &  16:39:35$\pm$1 &  0.32$\pm$0.01 &  103$\pm$7 &  - \\
18 &  16:46:44.7$\pm$0.7 &  0.37$\pm$0.01 &  38$\pm$3 &  - \\
18 &  17:09:21$\pm$4 &  0.33$\pm$0.04 &  117$\pm$22 &  - \\
20 &  12:14:27.5$\pm$0.3 &  0.54$\pm$0.02 &  17$\pm$1 &  - \\
20 &  14:05:33.7$\pm$0.7 &  0.42$\pm$0.02 &  43$\pm$3 &  - \\
21 &  11:12:39.2$\pm$0.8 &  0.36$\pm$0.01 &  45$\pm$3 &  - \\
21 &  13:08:27.6$\pm$0.1 &  3.1$\pm$0.2 &  3.0$\pm$0.5 &  0.001929$\pm$3e-06 \\
21 &  13:34:39.0$\pm$0.6 &  0.46$\pm$0.007 &  75$\pm$3 &  - \\
21 &  13:52:40.0$\pm$0.1 &  0.58$\pm$0.02 &  7.5$\pm$0.5 &  0.01$\pm$1e-05 \\
21 &  15:05:18.9$\pm$0.5 &  0.37$\pm$0.01 &  28$\pm$2 &  - \\
28 &  12:08:17.1$\pm$0.2 &  0.572$\pm$0.009 &  23.5$\pm$0.8 &  0.00717$\pm$1e-05 \\
28 &  12:22:10.20$\pm$0.08 &  1.49$\pm$0.09 &  2.6$\pm$0.3 &  - \\
28 &  14:16:47.63$\pm$0.08 &  0.86$\pm$0.03 &  4.2$\pm$0.3 &  - \\
28 &  14:21:38.1$\pm$0.9 &  0.42$\pm$0.02 &  44$\pm$4 &  - \\
29 &  13:05:30.1$\pm$0.7 &  1.03$\pm$0.02 &  88$\pm$4 &  - \\
29 &  13:17:07.0$\pm$0.1 &  0.61$\pm$0.08 &  2.0$\pm$0.4 &  - \\
29 &  16:38:00.9$\pm$0.5 &  0.33$\pm$0.01 &  33$\pm$2 &  - \\
30 &  18:18:27.7$\pm$0.7 &  0.92$\pm$0.01 &  138$\pm$4 &  - \\
31 &  11:01:25.3$\pm$0.6 &  0.54$\pm$0.01 &  55$\pm$3 &  - \\
31 &  11:24:07.5$\pm$0.7 &  0.23$\pm$0.02 &  19$\pm$2 &  - \\
31 &  11:33:53.6$\pm$0.2 &  0.44$\pm$0.02 &  8.7$\pm$0.8 &  0.002614$\pm$3e-06 \\
31 &  15:14:11$\pm$2 &  0.24$\pm$0.01 &  133$\pm$11 &  - \\
32 &  11:34:29.6$\pm$0.6 &  0.72$\pm$0.03 &  25$\pm$3 &  0.0103$\pm$2e-05 \\
32 &  11:51:17.6$\pm$0.3 &  0.39$\pm$0.01 &  24$\pm$1 &  - \\
32 &  13:51:43$\pm$2 &  1.44$\pm$0.06 &  183$\pm$12 &  - \\
32 &  13:51:57$\pm$1 &  0.14$\pm$0.03 &  12$\pm$5 &  - \\
33 &  12:10:53.3$\pm$0.3 &  1.16$\pm$0.03 &  29$\pm$1 &  - \\
33 &  13:02:24.7$\pm$0.2 &  0.52$\pm$0.05 &  4.3$\pm$0.7 &  - \\
33 &  15:19:31.08$\pm$0.06 &  1.11$\pm$0.04 &  3.1$\pm$0.2 &  - \\
33 &  16:37:57.2$\pm$0.5 &  0.37$\pm$0.03 &  11$\pm$2 &  - \\
34 &  11:53:33.3$\pm$0.6 &  0.54$\pm$0.01 &  64$\pm$3 &  - \\
34 &  12:14:27.5$\pm$0.3 &  0.5$\pm$0.03 &  10$\pm$1 &  - \\
34 &  12:19:54.47$\pm$0.06 &  1.06$\pm$0.03 &  4.1$\pm$0.2 &  - \\
34 &  12:24:42.1$\pm$0.9 &  0.43$\pm$0.02 &  42$\pm$4 &  - \\
34 &  12:30:12.2$\pm$0.1 &  1.96$\pm$0.04 &  13.6$\pm$0.6 &  0.00764$\pm$1e-05 \\
34 &  13:12:16.9$\pm$0.6 &  0.49$\pm$0.02 &  37$\pm$3 &  - \\
34 &  13:34:32.87$\pm$0.08 &  0.51$\pm$0.02 &  3.7$\pm$0.3 &  - \\
34 &  13:48:08.6$\pm$0.2 &  2.1$\pm$0.2 &  4.5$\pm$0.8 &  - \\
34 &  15:34:20.9$\pm$0.1 &  0.43$\pm$0.02 &  4.2$\pm$0.4 &  - \\
35 &  11:55:12.71$\pm$0.07 &  1.14$\pm$0.05 &  3.3$\pm$0.3 &  0.003678$\pm$3e-06 \\
35 &  12:05:00.9$\pm$0.7 &  0.33$\pm$0.01 &  33$\pm$3 &  - \\
35 &  14:15:51.4$\pm$0.6 &  0.29$\pm$0.01 &  32$\pm$2 &  - \\
36 &  10:17:11.7$\pm$0.3 &  1.2$\pm$0.09 &  9$\pm$1 &  - \\
36 &  12:11:01$\pm$1 &  0.395$\pm$0.009 &  109$\pm$5 &  - \\
36 &  12:43:54.4$\pm$0.1 &  1.49$\pm$0.06 &  6.5$\pm$0.5 &  - \\
37 &  12:11:19.68$\pm$0.09 &  1.18$\pm$0.03 &  7.9$\pm$0.4 &  - \\
37 &  14:03:13$\pm$1 &  0.31$\pm$0.02 &  41$\pm$5 &  - \\
37 &  14:13:46.9$\pm$0.1 &  1.14$\pm$0.05 &  5.5$\pm$0.5 &  0.035$\pm$0.002 \\
37 &  14:28:08.3$\pm$0.2 &  0.59$\pm$0.03 &  7.1$\pm$0.6 &  0.006537$\pm$5e-06 \\
38 &  12:04:31.4$\pm$0.3 &  0.723$\pm$0.01 &  40$\pm$1 &  - \\
38 &  12:30:43$\pm$1 &  0.43$\pm$0.01 &  121$\pm$6 &  - \\
38 &  13:09:44$\pm$2 &  0.59$\pm$0.02 &  175$\pm$10 &  0.007646$\pm$8e-06 \\
38 &  14:13:02$\pm$2 &  0.45$\pm$0.01 &  188$\pm$11 &  0.0585$\pm$0.0002 \\
38 &  15:17:50.2$\pm$0.1 &  0.64$\pm$0.04 &  3.8$\pm$0.4 &  - \\
38 &  16:15:45.3$\pm$0.5 &  0.3$\pm$0.01 &  26$\pm$2 &  - \\
39 &  11:02:53.8$\pm$0.2 &  1.26$\pm$0.04 &  10.9$\pm$0.7 &  - \\
39 &  12:09:53.63$\pm$0.07 &  1.6$\pm$0.1 &  2.2$\pm$0.3 &  - \\
39 &  12:26:22.9$\pm$0.3 &  0.42$\pm$0.02 &  11$\pm$1 &  - \\
39 &  14:03:08.2$\pm$0.7 &  0.48$\pm$0.01 &  57$\pm$3 &  0.081955$\pm$8e-06 \\
39 &  14:42:44.2$\pm$0.6 &  0.45$\pm$0.03 &  25$\pm$3 &  - \\
39 &  14:43:35.0$\pm$0.3 &  0.72$\pm$0.05 &  8$\pm$1 &  0.029783$\pm$9e-06 \\
41 &  11:39:03.9$\pm$0.4 &  0.54$\pm$0.02 &  17$\pm$1 &  - \\
41 &  12:14:02$\pm$1 &  0.35$\pm$0.01 &  70$\pm$5 &  - \\
41 &  13:18:14.0$\pm$0.1 &  0.65$\pm$0.01 &  11.2$\pm$0.5 &  0.009142$\pm$8e-06 \\
41 &  13:19:38.6$\pm$0.5 &  0.4$\pm$0.02 &  28$\pm$2 &  - \\
41 &  14:08:36$\pm$1 &  0.49$\pm$0.03 &  38$\pm$5 &  - \\
41 &  15:58:42.2$\pm$0.5 &  0.32$\pm$0.02 &  13$\pm$2 &  0.006002$\pm$5e-06 \\
44 &  10:52:21.7$\pm$0.3 &  0.3$\pm$0.01 &  12.3$\pm$0.9 &  0.01359$\pm$2e-05 \\
44 &  11:02:29.39$\pm$0.09 &  0.52$\pm$0.04 &  2.5$\pm$0.3 &  0.003106$\pm$2e-06 \\
44 &  12:49:22.8$\pm$0.3 &  0.84$\pm$0.02 &  27$\pm$1 &  - \\
44 &  13:49:15.9$\pm$0.2 &  0.62$\pm$0.03 &  10.4$\pm$0.9 &  0.01425$\pm$2e-05 \\
44 &  14:48:07$\pm$2 &  0.54$\pm$0.02 &  148$\pm$9 &  - \\
44 &  15:43:06.60$\pm$0.08 &  0.65$\pm$0.01 &  10.3$\pm$0.3 &  - \\
44 &  18:35:21$\pm$1 &  0.61$\pm$0.008 &  231$\pm$6 &  - \\
45 &  10:39:43.2$\pm$0.2 &  0.46$\pm$0.03 &  7.6$\pm$1.0 &  0.004318$\pm$4e-06 \\
45 &  10:51:49.3$\pm$0.4 &  0.485$\pm$0.01 &  40$\pm$2 &  0.00466$\pm$6e-06 \\
45 &  11:33:16.9$\pm$0.1 &  0.46$\pm$0.02 &  6.5$\pm$0.6 &  0.023436$\pm$8e-06 \\
45 &  11:57:10.4$\pm$0.5 &  0.57$\pm$0.03 &  19$\pm$2 &  0.006427$\pm$6e-06 \\
45 &  12:57:56.1$\pm$0.3 &  1.14$\pm$0.07 &  12$\pm$1 &  - \\
45 &  12:58:57.3$\pm$0.4 &  0.53$\pm$0.03 &  13$\pm$2 &  0.03615$\pm$7e-05 \\
45 &  13:36:29.52$\pm$0.1 &  0.65$\pm$0.08 &  2.0$\pm$0.4 &  - \\
45 &  14:52:28.5$\pm$0.4 &  0.44$\pm$0.01 &  30$\pm$2 &  - \\
45 &  19:11:31.9$\pm$0.7 &  0.379$\pm$0.009 &  59$\pm$3 &  - \\
46 &  12:09:42.1$\pm$0.2 &  0.43$\pm$0.03 &  4.7$\pm$0.6 &  - \\
46 &  13:10:42$\pm$1 &  0.6$\pm$0.03 &  41$\pm$4 &  - \\
46 &  13:12:26.28$\pm$0.06 &  0.65$\pm$0.03 &  2.4$\pm$0.2 &  - \\
46 &  14:48:43$\pm$2 &  0.56$\pm$0.02 &  186$\pm$11 &  0.004882$\pm$9e-06 \\
46 &  16:14:36.6$\pm$0.8 &  0.33$\pm$0.05 &  10$\pm$3 &  - \\
47 &  11:13:02$\pm$1 &  0.36$\pm$0.01 &  92$\pm$6 &  - \\
47 &  12:26:50.6$\pm$0.6 &  0.24$\pm$0.01 &  23$\pm$3 &  0.003309$\pm$4e-06 \\
47 &  12:39:15.8$\pm$0.5 &  0.27$\pm$0.01 &  20$\pm$2 &  - \\
47 &  12:48:36.3$\pm$0.4 &  0.39$\pm$0.02 &  16$\pm$2 &  - \\
47 &  12:49:10$\pm$2 &  0.47$\pm$0.02 &  106$\pm$8 &  0.006732$\pm$8e-06 \\
47 &  14:10:16.3$\pm$0.2 &  0.46$\pm$0.01 &  18.8$\pm$0.9 &  0.01703$\pm$1e-05 \\
47 &  15:43:02.2$\pm$0.2 &  1.1$\pm$0.2 &  2.9$\pm$0.8 &  - \\
47 &  15:52:47.2$\pm$0.7 &  0.75$\pm$0.02 &  90$\pm$4 &  - \\
47 &  16:03:08.9$\pm$0.9 &  0.34$\pm$0.02 &  32$\pm$4 &  - \\
47 &  19:20:48.8$\pm$0.3 &  0.51$\pm$0.03 &  9$\pm$1 &  - \\
48 &  12:53:03$\pm$4 &  0.39$\pm$0.04 &  219$\pm$27 &  - \\
48 &  15:19:15$\pm$2 &  0.39$\pm$0.01 &  102$\pm$8 &  - \\
48 &  15:31:36$\pm$2 &  0.39$\pm$0.02 &  119$\pm$12 &  - \\
48 &  18:28:51$\pm$1 &  0.466$\pm$0.009 &  163$\pm$7 &  - \\
49 &  12:30:35.28$\pm$0.07 &  0.85$\pm$0.02 &  6.6$\pm$0.3 &  - \\
49 &  13:18:36.8$\pm$0.7 &  0.37$\pm$0.01 &  42$\pm$3 &  - \\
49 &  15:26:18$\pm$1 &  0.36$\pm$0.03 &  31$\pm$5 &  - \\
49 &  15:33:17.2$\pm$0.3 &  0.46$\pm$0.01 &  22$\pm$1 &  - \\
49 &  17:00:45.7$\pm$0.9 &  0.73$\pm$0.01 &  116$\pm$4 &  - \\
50 &  11:19:46.1$\pm$0.9 &  0.8$\pm$0.02 &  71$\pm$4 &  - \\
50 &  11:20:08.5$\pm$0.5 &  0.33$\pm$0.03 &  18$\pm$3 &  - \\
50 &  11:24:50.3$\pm$0.4 &  0.55$\pm$0.01 &  31$\pm$2 &  - \\
50 &  11:32:53.1$\pm$0.2 &  0.61$\pm$0.02 &  12.4$\pm$0.7 &  0.022157$\pm$5e-06 \\
50 &  11:50:12.8$\pm$0.1 &  1.09$\pm$0.03 &  8.9$\pm$0.4 &  - \\
50 &  12:10:51$\pm$3 &  0.5$\pm$0.02 &  123$\pm$12 &  - \\
50 &  13:30:48.6$\pm$0.2 &  0.68$\pm$0.05 &  4.7$\pm$0.5 &  - \\
50 &  15:23:47.39$\pm$0.09 &  1.4$\pm$0.05 &  5.9$\pm$0.4 &  - \\
50 &  16:31:26.7$\pm$0.5 &  0.89$\pm$0.02 &  51$\pm$2 &  - \\
51 &  11:54:35.6$\pm$0.5 &  0.28$\pm$0.02 &  18$\pm$2 &  0.01686$\pm$2e-05 \\
51 &  12:15:08.6$\pm$0.2 &  0.74$\pm$0.08 &  4.1$\pm$0.8 &  - \\
52 &  12:53:24.1$\pm$0.2 &  0.478$\pm$0.009 &  17.5$\pm$0.6 &  - \\
52 &  13:08:02.7$\pm$0.4 &  0.37$\pm$0.02 &  21$\pm$2 &  0.01602$\pm$2e-05 \\
52 &  13:21:52.55$\pm$0.1 &  0.43$\pm$0.06 &  2.0$\pm$0.5 &  - \\
52 &  14:07:33.9$\pm$0.5 &  0.32$\pm$0.01 &  29$\pm$2 &  - \\
52 &  15:52:08.3$\pm$0.1 &  0.51$\pm$0.02 &  8.4$\pm$0.6 &  - \\
53 &  11:05:57.11$\pm$0.08 &  0.43$\pm$0.03 &  2.3$\pm$0.2 &  - \\
53 &  11:56:24$\pm$3 &  0.36$\pm$0.01 &  184$\pm$11 &  0.004066$\pm$8e-06 \\
53 &  12:10:54$\pm$1 &  0.21$\pm$0.02 &  30$\pm$4 &  - \\
53 &  12:55:47.6$\pm$0.2 &  0.46$\pm$0.02 &  11.7$\pm$0.8 &  0.01336$\pm$2e-05 \\
53 &  15:03:49.3$\pm$0.8 &  0.35$\pm$0.01 &  40$\pm$3 &  - \\
54 &  11:29:29.04$\pm$0.06 &  0.98$\pm$0.02 &  5.1$\pm$0.2 &  - \\
54 &  13:58:28$\pm$1 &  0.29$\pm$0.01 &  101$\pm$8 &  0.0386$\pm$0.0001 \\
54 &  15:36:35.2$\pm$0.5 &  0.46$\pm$0.02 &  21$\pm$2 &  - \\
55 &  12:14:22.5$\pm$0.4 &  1.56$\pm$0.05 &  25$\pm$2 &  - \\
55 &  12:22:21.3$\pm$0.1 &  0.92$\pm$0.02 &  9.9$\pm$0.5 &  - \\
55 &  12:27:51.8$\pm$1.0 &  0.49$\pm$0.01 &  98$\pm$4 &  - \\
55 &  12:38:27.9$\pm$0.1 &  0.68$\pm$0.01 &  14.5$\pm$0.6 &  0.01891$\pm$1e-05 \\
55 &  12:53:00$\pm$1 &  0.287$\pm$0.009 &  80$\pm$5 &  - \\
55 &  14:31:22.4$\pm$0.6 &  0.489$\pm$0.007 &  83$\pm$3 &  - \\
55 &  15:22:30.7$\pm$0.4 &  0.464$\pm$0.009 &  45$\pm$2 &  0.0725$\pm$0.0001 \\
56 &  10:19:20.99$\pm$0.08 &  0.45$\pm$0.02 &  3.3$\pm$0.2 &  - \\
56 &  11:32:18.2$\pm$0.5 &  0.45$\pm$0.02 &  31$\pm$2 &  - \\
56 &  12:05:13.91$\pm$0.06 &  0.82$\pm$0.02 &  5.7$\pm$0.2 &  - \\
56 &  12:35:44.5$\pm$0.3 &  1.5$\pm$0.1 &  8$\pm$1 &  0.002214$\pm$6e-06 \\
56 &  13:15:29.1$\pm$0.4 &  0.33$\pm$0.01 &  19$\pm$2 &  - \\
56 &  13:28:18.8$\pm$0.3 &  0.61$\pm$0.02 &  21$\pm$1 &  - \\
56 &  14:38:12.4$\pm$0.2 &  0.53$\pm$0.01 &  19.9$\pm$0.8 &  - \\
56 &  14:54:32.0$\pm$0.3 &  0.98$\pm$0.06 &  8.8$\pm$1.0 &  - \\
57 &  10:58:31.0$\pm$0.2 &  2.8$\pm$0.2 &  6.7$\pm$0.9 &  - \\
57 &  12:38:51.1$\pm$0.4 &  2.9$\pm$0.1 &  14$\pm$2 &  - \\
57 &  12:41:53.1$\pm$0.2 &  1.01$\pm$0.03 &  11.3$\pm$0.7 &  - \\
58 &  16:18:47.8$\pm$0.3 &  0.43$\pm$0.02 &  11.2$\pm$0.9 &  - \\
59 &  12:13:04.4$\pm$0.4 &  0.334$\pm$0.009 &  26$\pm$1 &  0.003362$\pm$8e-06 \\
59 &  12:55:14.1$\pm$0.4 &  0.49$\pm$0.01 &  29$\pm$1 &  - \\
59 &  13:07:29.28$\pm$0.05 &  1.88$\pm$0.06 &  3.2$\pm$0.2 &  - \\
59 &  13:18:22.68$\pm$0.05 &  0.78$\pm$0.01 &  5.1$\pm$0.2 &  - \\
59 &  13:42:25$\pm$1 &  0.334$\pm$0.007 &  106$\pm$5 &  - \\
59 &  14:07:57.0$\pm$0.1 &  0.8$\pm$0.03 &  6.1$\pm$0.4 &  0.005892$\pm$4e-06 \\
59 &  15:37:02.9$\pm$0.7 &  0.64$\pm$0.01 &  89$\pm$4 &  - \\
60 &  12:19:35.7$\pm$0.9 &  0.35$\pm$0.04 &  17$\pm$3 &  - \\
60 &  12:52:41$\pm$1 &  0.28$\pm$0.02 &  40$\pm$5 &  0.007498$\pm$7e-06 \\
60 &  13:00:55.07$\pm$0.04 &  0.74$\pm$0.02 &  2.0$\pm$0.1 &  - \\
61 &  10:52:59.1$\pm$0.6 &  0.44$\pm$0.02 &  28$\pm$2 &  - \\
61 &  11:15:51$\pm$2 &  0.35$\pm$0.01 &  151$\pm$10 &  - \\
61 &  12:11:33.3$\pm$0.3 &  0.4$\pm$0.01 &  19$\pm$1 &  - \\
61 &  12:39:43.5$\pm$0.3 &  1.42$\pm$0.05 &  18$\pm$1 &  - \\
61 &  13:14:30.8$\pm$0.1 &  1.8$\pm$0.1 &  4.9$\pm$0.6 &  - \\
61 &  14:06:03.2$\pm$0.7 &  0.43$\pm$0.01 &  47$\pm$3 &  - \\
61 &  14:53:56.3$\pm$0.4 &  0.34$\pm$0.01 &  19$\pm$1 &  - \\
61 &  14:58:58.4$\pm$0.4 &  0.34$\pm$0.01 &  25$\pm$1 &  - \\
62 &  12:27:13.31$\pm$0.07 &  0.62$\pm$0.03 &  3.3$\pm$0.3 &  - \\
62 &  12:31:09.83$\pm$0.09 &  0.72$\pm$0.04 &  3.0$\pm$0.3 &  0.012654$\pm$1e-05 \\
62 &  12:45:22.3$\pm$0.5 &  0.78$\pm$0.03 &  26$\pm$2 &  0.002037$\pm$5e-06 \\
63 &  11:02:00.6$\pm$0.6 &  0.52$\pm$0.01 &  53$\pm$3 &  - \\
63 &  11:53:52$\pm$1 &  0.56$\pm$0.01 &  203$\pm$7 &  - \\
63 &  13:43:49$\pm$1 &  0.592$\pm$0.01 &  167$\pm$6 &  - \\
63 &  15:51:46.0$\pm$0.3 &  0.46$\pm$0.02 &  19$\pm$1 &  - \\
63 &  16:30:29.5$\pm$0.3 &  0.48$\pm$0.02 &  13.1$\pm$1.0 &  - \\
64 &  12:03:05.0$\pm$0.9 &  0.35$\pm$0.01 &  46$\pm$3 &  - \\
64 &  12:20:08.5$\pm$0.3 &  1.28$\pm$0.03 &  25$\pm$1 &  - \\
64 &  12:52:59.1$\pm$0.3 &  1.15$\pm$0.03 &  21$\pm$1 &  0.05887$\pm$1e-05 \\
64 &  12:55:37.9$\pm$0.4 &  0.36$\pm$0.02 &  17$\pm$2 &  - \\
64 &  13:22:07.3$\pm$0.8 &  0.663$\pm$0.007 &  197$\pm$4 &  - \\
64 &  13:22:21$\pm$1 &  0.15$\pm$0.01 &  39$\pm$6 &  - \\
65 &  11:11:07.79$\pm$0.08 &  0.97$\pm$0.04 &  4.0$\pm$0.3 &  - \\
65 &  11:47:28$\pm$1 &  0.33$\pm$0.02 &  64$\pm$7 &  - \\
65 &  12:21:53.64$\pm$0.09 &  0.62$\pm$0.03 &  3.8$\pm$0.3 &  - \\
65 &  12:29:34.0$\pm$0.3 &  2.4$\pm$0.1 &  16$\pm$2 &  - \\
65 &  12:56:14.64$\pm$0.04 &  1.56$\pm$0.02 &  4.7$\pm$0.1 &  - \\
65 &  12:57:23.4$\pm$0.1 &  1.4$\pm$0.2 &  2.5$\pm$0.5 &  0.010328$\pm$9e-06 \\
65 &  13:39:58.3$\pm$0.5 &  0.34$\pm$0.03 &  12$\pm$2 &  0.08073$\pm$2e-05 \\
65 &  14:19:06.2$\pm$0.2 &  0.65$\pm$0.02 &  15.1$\pm$0.9 &  - \\
65 &  14:35:22$\pm$3 &  0.46$\pm$0.06 &  139$\pm$19 &  - \\
65 &  15:17:11.4$\pm$0.3 &  0.47$\pm$0.01 &  22$\pm$1 &  - \\
66 &  12:43:59$\pm$1 &  0.339$\pm$0.009 &  95$\pm$6 &  - \\
66 &  12:49:56.9$\pm$0.3 &  0.38$\pm$0.01 &  18.3$\pm$1.0 &  0.01362$\pm$1e-05 \\
66 &  13:41:19.3$\pm$0.3 &  0.63$\pm$0.04 &  8.3$\pm$1.0 &  - \\
67 &  11:16:10$\pm$1 &  0.85$\pm$0.02 &  111$\pm$5 &  - \\
67 &  11:21:31$\pm$3 &  0.35$\pm$0.02 &  157$\pm$16 &  - \\
67 &  11:26:07$\pm$2 &  0.39$\pm$0.02 &  65$\pm$7 &  - \\
67 &  12:32:09$\pm$1 &  0.44$\pm$0.01 &  131$\pm$7 &  - \\
67 &  12:48:18.3$\pm$0.1 &  0.47$\pm$0.02 &  7.4$\pm$0.4 &  0.003967$\pm$4e-06 \\
67 &  13:56:54$\pm$3 &  0.25$\pm$0.02 &  110$\pm$16 &  - \\
68 &  08:50:35.9$\pm$0.2 &  0.23$\pm$0.01 &  8.4$\pm$0.8 &  - \\
68 &  11:22:26$\pm$2 &  0.44$\pm$0.01 &  228$\pm$9 &  - \\
68 &  12:03:29$\pm$2 &  0.25$\pm$0.01 &  112$\pm$12 &  - \\
68 &  12:57:34.5$\pm$0.8 &  0.39$\pm$0.01 &  54$\pm$3 &  - \\
68 &  14:08:33.62$\pm$0.09 &  0.94$\pm$0.02 &  7.3$\pm$0.4 &  - \\
68 &  14:09:57.2$\pm$0.8 &  0.43$\pm$0.02 &  48$\pm$5 &  0.007514$\pm$5e-06 \\
68 &  14:24:38.5$\pm$0.2 &  0.78$\pm$0.01 &  23.6$\pm$0.7 &  - \\
68 &  14:30:35.2$\pm$0.2 &  0.42$\pm$0.01 &  16.1$\pm$0.8 &  - \\
69 &  13:07:32.8$\pm$0.9 &  0.46$\pm$0.02 &  51$\pm$4 &  - \\
69 &  13:57:49.6$\pm$0.1 &  0.91$\pm$0.02 &  12.4$\pm$0.5 &  - \\
69 &  15:08:30.8$\pm$0.2 &  0.73$\pm$0.05 &  6.5$\pm$0.8 &  0.009759$\pm$8e-06 \\
69 &  15:49:45.1$\pm$0.3 &  0.592$\pm$0.01 &  34$\pm$1 &  - \\
70 &  10:25:11.99$\pm$0.09 &  1.03$\pm$0.04 &  5.1$\pm$0.4 &  - \\
71 &  11:38:00.2$\pm$0.4 &  0.54$\pm$0.02 &  20$\pm$2 &  - \\
71 &  11:55:10.1$\pm$0.8 &  0.6$\pm$0.02 &  50$\pm$3 &  - \\
71 &  12:26:44$\pm$1 &  0.287$\pm$0.007 &  110$\pm$5 &  - \\
71 &  12:57:52.9$\pm$0.2 &  0.71$\pm$0.03 &  8.7$\pm$0.7 &  - \\
71 &  14:31:10.9$\pm$0.1 &  0.87$\pm$0.04 &  4.4$\pm$0.4 &  - \\
71 &  15:13:28.5$\pm$1.0 &  0.392$\pm$0.008 &  110$\pm$5 &  0.02474$\pm$2e-05 \\
72 &  11:20:52.0$\pm$0.8 &  0.262$\pm$0.009 &  50$\pm$3 &  - \\
72 &  11:26:28$\pm$1 &  0.427$\pm$0.009 &  141$\pm$6 &  - \\
72 &  12:28:39$\pm$1 &  0.64$\pm$0.02 &  127$\pm$6 &  - \\
72 &  12:29:05.6$\pm$0.6 &  0.33$\pm$0.02 &  41$\pm$4 &  0.01321$\pm$1e-05 \\
72 &  12:48:41.0$\pm$0.8 &  0.33$\pm$0.01 &  49$\pm$3 &  0.003136$\pm$7e-06 \\
72 &  12:59:02.0$\pm$0.7 &  0.306$\pm$0.006 &  72$\pm$3 &  - \\
72 &  14:05:13.5$\pm$0.9 &  0.291$\pm$0.007 &  78$\pm$4 &  - \\
72 &  14:32:41.9$\pm$0.9 &  1.09$\pm$0.06 &  38$\pm$4 &  0.008483$\pm$7e-06 \\
72 &  15:33:27.3$\pm$0.2 &  0.41$\pm$0.01 &  16.0$\pm$0.8 &  - \\
73 &  13:19:19.56$\pm$0.1 &  0.75$\pm$0.03 &  5.8$\pm$0.4 &  0.01496$\pm$2e-05 \\
73 &  14:34:33$\pm$1 &  0.32$\pm$0.01 &  78$\pm$6 &  - \\
74 &  09:54:41.7$\pm$0.4 &  0.23$\pm$0.02 &  12$\pm$1 &  0.003538$\pm$4e-06 \\
74 &  13:05:16.0$\pm$0.2 &  0.34$\pm$0.02 &  10.0$\pm$0.9 &  - \\
74 &  13:17:49.2$\pm$0.1 &  0.57$\pm$0.05 &  2.9$\pm$0.5 &  - \\
74 &  13:39:46$\pm$2 &  0.32$\pm$0.01 &  117$\pm$9 &  - \\
75 &  11:08:53.1$\pm$0.2 &  0.367$\pm$0.009 &  12.8$\pm$0.6 &  - \\
75 &  12:00:05$\pm$3 &  0.49$\pm$0.04 &  217$\pm$20 &  0.00478$\pm$1e-05 \\
75 &  12:23:07.0$\pm$0.4 &  0.41$\pm$0.02 &  14$\pm$1 &  0.01721$\pm$2e-05 \\
75 &  14:35:53$\pm$1 &  0.574$\pm$0.009 &  140$\pm$5 &  - \\
76 &  10:25:14.87$\pm$0.06 &  0.74$\pm$0.01 &  8.1$\pm$0.3 &  - \\
76 &  10:46:54.1$\pm$0.1 &  0.63$\pm$0.06 &  2.0$\pm$0.3 &  - \\
76 &  11:53:22.9$\pm$0.7 &  0.41$\pm$0.01 &  50$\pm$3 &  - \\
76 &  14:12:16.9$\pm$0.8 &  0.3$\pm$0.03 &  18$\pm$3 &  - \\
76 &  14:17:02.0$\pm$0.7 &  0.36$\pm$0.01 &  37$\pm$3 &  - \\
77 &  11:36:19.7$\pm$0.9 &  0.49$\pm$0.02 &  71$\pm$5 &  0.003699$\pm$7e-06 \\
77 &  13:17:52.4$\pm$0.7 &  0.25$\pm$0.02 &  15$\pm$3 &  0.005849$\pm$7e-06 \\
77 &  13:41:12.4$\pm$0.6 &  0.53$\pm$0.01 &  47$\pm$3 &  0.034374$\pm$8e-06 \\
78 &  10:43:09.1$\pm$0.3 &  0.41$\pm$0.01 &  17$\pm$1 &  - \\
78 &  12:00:12.6$\pm$0.4 &  1.9$\pm$0.1 &  14$\pm$2 &  0.01814$\pm$2e-05 \\
78 &  12:35:56.0$\pm$0.2 &  0.56$\pm$0.01 &  21$\pm$1 &  - \\
78 &  14:49:45$\pm$1 &  0.41$\pm$0.009 &  126$\pm$6 &  0.009987$\pm$7e-06 \\
79 &  10:54:59.3$\pm$0.1 &  0.61$\pm$0.01 &  10.1$\pm$0.5 &  - \\
79 &  11:46:06.9$\pm$0.6 &  0.38$\pm$0.03 &  20$\pm$3 &  - \\
79 &  12:51:23.7$\pm$0.2 &  0.7$\pm$0.02 &  10.8$\pm$0.7 &  0.015443$\pm$8e-06 \\
79 &  13:47:20$\pm$1 &  0.364$\pm$0.01 &  88$\pm$5 &  - \\
80 &  12:05:38$\pm$1 &  0.36$\pm$0.02 &  46$\pm$6 &  - \\
80 &  12:30:50.7$\pm$0.3 &  0.34$\pm$0.01 &  19$\pm$1 &  - \\
80 &  12:48:07$\pm$1 &  0.383$\pm$0.009 &  105$\pm$5 &  - \\
80 &  14:44:59.6$\pm$0.2 &  1.6$\pm$0.1 &  4.8$\pm$0.8 &  - \\
80 &  16:04:35.3$\pm$0.1 &  0.66$\pm$0.02 &  12.5$\pm$0.6 &  - \\
81 &  09:53:54.8$\pm$0.1 &  0.4$\pm$0.01 &  7.0$\pm$0.5 &  - \\
81 &  11:31:58.4$\pm$0.1 &  0.54$\pm$0.01 &  10.3$\pm$0.5 &  - \\
81 &  11:45:51.1$\pm$0.3 &  0.56$\pm$0.02 &  13$\pm$1 &  0.004808$\pm$7e-06 \\
81 &  12:56:14.99$\pm$0.05 &  1.38$\pm$0.02 &  6.2$\pm$0.2 &  0.034304$\pm$6e-06 \\
81 &  14:53:54.24$\pm$0.08 &  0.55$\pm$0.06 &  2.0$\pm$0.3 &  - \\
82 &  10:37:25.3$\pm$0.1 &  0.72$\pm$0.06 &  3.0$\pm$0.4 &  - \\
82 &  12:04:40.0$\pm$0.6 &  5.7$\pm$0.9 &  7$\pm$2 &  - \\
82 &  14:43:23.1$\pm$0.2 &  0.33$\pm$0.02 &  9.6$\pm$0.9 &  - \\
83 &  11:19:42.9$\pm$0.1 &  0.59$\pm$0.01 &  11.2$\pm$0.5 &  0.009447$\pm$6e-06 \\
83 &  11:54:38.5$\pm$0.2 &  0.87$\pm$0.05 &  6.2$\pm$0.7 &  0.01996$\pm$2e-05 \\
83 &  11:55:50.51$\pm$0.07 &  0.75$\pm$0.02 &  5.8$\pm$0.3 &  - \\
83 &  14:51:07$\pm$2 &  0.24$\pm$0.01 &  92$\pm$9 &  - \\
84 &  12:33:37.4$\pm$0.7 &  0.44$\pm$0.02 &  32$\pm$3 &  - \\
84 &  15:53:17.1$\pm$0.1 &  0.66$\pm$0.03 &  5.4$\pm$0.4 &  0.0061$\pm$9e-05 \\
85 &  11:23:38.0$\pm$0.8 &  0.32$\pm$0.02 &  33$\pm$3 &  - \\
85 &  11:44:22.9$\pm$0.3 &  1.62$\pm$0.04 &  28$\pm$1 &  0.02113$\pm$3e-05 \\
85 &  11:53:02.0$\pm$0.4 &  0.46$\pm$0.02 &  19$\pm$1 &  6e-06$\pm$2e-06 \\
85 &  13:43:19.9$\pm$0.9 &  0.42$\pm$0.01 &  84$\pm$5 &  - \\
85 &  13:55:49.8$\pm$0.1 &  0.58$\pm$0.02 &  9.0$\pm$0.6 &  0.007595$\pm$9e-06 \\
85 &  15:23:41.9$\pm$0.4 &  0.44$\pm$0.02 &  16$\pm$1 &  - \\
85 &  15:35:00.2$\pm$0.2 &  2.3$\pm$0.2 &  3.4$\pm$0.6 &  0.01833$\pm$8e-05 \\
86 &  10:40:14.16$\pm$0.06 &  0.53$\pm$0.01 &  4.8$\pm$0.2 &  - \\
86 &  12:02:08.5$\pm$0.4 &  0.7$\pm$0.01 &  58$\pm$2 &  - \\
86 &  12:34:01.55$\pm$0.04 &  0.61$\pm$0.01 &  4.1$\pm$0.1 &  - \\
86 &  14:23:45.2$\pm$0.2 &  0.43$\pm$0.01 &  13.7$\pm$0.8 &  - \\
87 &  13:02:46.6$\pm$0.4 &  0.73$\pm$0.02 &  35$\pm$2 &  - \\
87 &  13:11:54$\pm$1 &  0.31$\pm$0.02 &  46$\pm$5 &  - \\
87 &  13:18:45.3$\pm$0.5 &  0.326$\pm$0.006 &  53$\pm$2 &  - \\
87 &  13:51:23$\pm$1 &  0.62$\pm$0.01 &  141$\pm$6 &  - \\
88 &  12:04:16.3$\pm$0.1 &  0.9$\pm$0.02 &  11.2$\pm$0.6 &  0.02274$\pm$3e-05 \\
89 &  11:21:25.5$\pm$0.2 &  1.41$\pm$0.03 &  21.5$\pm$0.9 &  - \\
89 &  13:27:24$\pm$3 &  0.36$\pm$0.01 &  202$\pm$16 &  - \\
89 &  13:54:46.8$\pm$0.2 &  0.54$\pm$0.02 &  8.3$\pm$0.7 &  - \\
89 &  14:57:56.52$\pm$0.03 &  2.65$\pm$0.05 &  3.5$\pm$0.1 &  - \\
\enddata


\tablecomments{For encounters without statistically meaningful RDS excursions, ${\rm max} |F - \overline{F}|/\overline{F}$ is not given.}


\end{deluxetable}
\label{tbl:vortex_fit_params}

\begin{acknowledgments}
This study benefited from conversations with Ryan Battin, Justin Crevier, Lori Fenton, Ralph Lorenz, and Michelle Szurgot. It also benefited from thoughtful and thorough feedback from two anonymous referees. This research was supported by a grant from NASA's Solar System Workings program, NNH17ZDA001N-SSW.
\end{acknowledgments}

\software{matplotlib \citep{Hunter:2007}, numpy \citep{harris2020array}, scipy \citep{2020SciPy-NMeth}, Astropy \citep{astropy:2013, astropy:2018}}

\bibliography{sample631}{}
\bibliographystyle{aasjournal}



\end{document}